\documentclass[11pt]{iopart}

\expandafter\let\csname equation*\endcsname\relax
\expandafter\let\csname endequation*\endcsname\relax

\newcommand*{\affaddr}[1]{#1} 
\newcommand*{\affmark}[1][*]{\textsuperscript{#1}}

\usepackage{tabularx} 
\usepackage{amsmath}  
\usepackage{graphicx} 
\usepackage[margin=1in,letterpaper]{geometry} 
\usepackage{cite} %

\usepackage{lipsum}

\usepackage{hyperref}
\usepackage{breqn}
\usepackage{mathtools}

\usepackage{breqn}
\usepackage{amssymb}
\usepackage{subcaption}
\usepackage[toc,page]{appendix}

\hypersetup{
     colorlinks   = true,
     citecolor    = blue,
     linkcolor    = blue}

\DeclareMathOperator\erfc{erfc}

\begin{document}

\title[Sciortino \emph{et al.} - Inference of Experimental Radial Impurity Transport]{Inference of Experimental Radial Impurity Transport on Alcator C-Mod: Bayesian Parameter Estimation and Model Selection.}

\author{ F. Sciortino\affmark[1]\footnote{Email: sciortino@psfc.mit.edu}, N.T. Howard\affmark[1], E.S. Marmar\affmark[1], T. Odstrcil\affmark[1], N.M. Cao\affmark[1], R. Dux\affmark[2], A.E. Hubbard\affmark[1], J.W. Hughes\affmark[1], J.H. Irby\affmark[1], Y. Marzouk\affmark[1], L.M. Milanese\affmark[1], M.L. Reinke\affmark[3], J.E. Rice\affmark[1], P. Rodriguez-Fernandez\affmark[1]}
\affaddr{\affmark[1]Massachusetts Institute of Technology, Cambridge, MA, 02139, USA}\\
\affaddr{\affmark[2]Max Planck Institute - Institute for Plasma Physics, Garching, Germany}\\
\affaddr{\affmark[3]Oak Ridge National Laboratory, Oak Ridge, TN 37831, USA}\\

\begin{abstract}
We present a fully Bayesian approach for the inference of radial profiles of impurity transport coefficients and compare its results to neoclassical, gyrofluid and gyrokinetic modeling. Using nested sampling, the Bayesian Impurity Transport InferencE (\texttt{BITE}) framework can handle complex parameter spaces with multiple possible solutions, offering great advantages in interpretative power and reliability with respect to previously demonstrated methods. \texttt{BITE} employs a forward model based on the \texttt{pySTRAHL} package, built on the success of the well-known \texttt{STRAHL} code [Dux, IPP Report, 2004], to simulate impurity transport in magnetically-confined plasmas. In this paper, we focus on calcium (Ca, Z=20) Laser Blow-Off injections into Alcator C-Mod plasmas. Multiple Ca atomic lines are diagnosed via high-resolution X-ray Imaging Crystal Spectroscopy and Vacuum Ultra-Violet measurements. We analyze a sawtoothing I-mode discharge for which neoclassical and turbulent (quasilinear and nonlinear) predictions are also obtained. We find good agreement in diffusion across the entire radial extent, while turbulent convection and density profile peaking are estimated to be larger in experiment than suggested by theory. Efforts and challenges associated with the inference of experimental pedestal impurity transport are discussed. 
\end{abstract}

\section{Introduction} \label{sec:intro}
Stable high-performance tokamak operation requires complex trade-offs to maintain acceptable impurity conditions. Core-edge integration demands minimization of core radiation and dilution\cite{Putterich2019DeterminationFactors} while also sustaining a detached divertor regime, envisioned to only be possible via puffing of impurities in the edge\cite{Reinke2017ExpandingDivertors}. Choices of wall materials are also known to affect plasma fueling, pedestal formation and high-confinement (H-mode) operation\cite{Dunne2017ImpactELMs}. Developing and validating accurate models for impurity transport and atomic data are paramount to make predictions in a whole-device perspective. 

Historically, transport model validation for fusion plasmas has mostly focused on heat transport\cite{Holland2016ValidationTransport, Greenwald2010VerificationFusion}. Accurate estimation of experimental particle transport coefficients has proven to be more demanding, although significant progress has been made, particularly on understanding the origin of electron density profile peaking \cite{Angioni2003DensityPlasmas, Angioni2009ParticleExperiment, Greenwald2007DensityScalings, Angioni2014TungstenModeling, Tala2019DensityTransport}. Detailed validation of impurity transport theory is more challenging, both from an experimental and a theoretical perspective. 

Since transport codes tend to make predictions for \emph{fluxes}, validation efforts would ideally compare \emph{fluxes} between experiments and theoretical predictions. This, however, would be a formidable task for impurity transport, to the required level of accuracy, since knowledge of sources, atomic rates and diagnostic calibrations each present significant challenges. We thus resort to transport coefficients, diffusion ($D$) and convection ($v$), to separate the particle flux ($\Gamma$) via $\Gamma \coloneqq -D \ \nabla n + v \ n$. $D$ and $v$ are more effective metrics to compare experiment and theory without over-reliance on experimental details. In quasi-steady conditions with no impurity core sources, we may expect $\Gamma=0$, which allows one to relate particle density gradients to the $v/D$ ratio via $v/D=\nabla n/n$. It is only in the presence of time-dependent dynamics that $D$ and $v$ may be experimentally separated, offering much stronger constraints for transport model validation. For low-Z ions that may be assumed to be fully-ionized, core transport coefficients may be estimated using relatively simple flux-gradient methods~\cite{Wade1995HeliumDIII-D}, which however do not generalize to higher-Z impurities for which atomic physics plays a key role. In this work, we make use of transient impurity injections produced via the Laser Blow-Off (LBO) technique\cite{Marmar1975SystemPlasmas}.


The inference of $D$ and $v$ relies on an iterative process in which predictions of charge state densities and emissivity profiles are computed for given sets of parameters choices, attempting to match experimental observations. For each choice of parameters, a ``goodness-of-fit'' metric is calculated in order to identify parameters that most closely match reality. This is referred to as an \emph{inverse problem}, as opposed to the operation of computing a signal prediction for a given choice of model and parameters (\emph{forward modeling}). Ideally, we would like to solve an inverse problem with the lowest possible number of free parameters (inputs to the forward model) while still reproducing experimental behavior to the highest possible degree of fidelity. 

Efforts to estimate radial profiles of $D$ and $v$ have been reported from all major devices, including ASDEX-Upgrade \cite{Dux2003ImpurityUpgrade, Dux2003InfluenceUpgrade, Putterich2011ELMUpgrade,Bruhn2020Erratum:10.1088/1361-6587/aac870}, DIII-D \cite{Grierson2015ImpurityDIII-D,Odstrcil_2020}, JET\cite{Dux2004ImpurityJET,Giroud2007MethodJET}, Tore Supra\cite{Parisot2008ExperimentalPlasma, Villegas2014ExperimentalSupra}, TCV \cite{Scavino2003EffectsTCV}, HL-2A \cite{Cui2018StudyDepositions}, W7-X\cite{Geiger2019ObservationIron}, MAST\cite{Henderson2015ChargeMAST}, NSTX~\cite{Delgado-Aparicio2009ImpurityPlasmas} and Alcator C-Mod \cite{Rice2000ImpurityPlasmas, Pedersen2000RadialC-Mod, Rice2007ImpurityPlasmas,Howard2012QuantitativePlasma, Chilenski2018EfficientExperiments}. In several studies, gas puffing or supersonic pellets were used to inject small amounts of impurities; both of these methods may incur difficulties when modeling the impurity source. On the other hand, the Laser Blow-Off (LBO) technique\cite{Marmar1975SystemPlasmas} allows the injection of non-recycling impurities via a time-resolved and non-perturbative laser ablation, whose source function can be more easily characterized. 


On Alcator C-Mod, the inference of experimental impurity transport coefficients was previously attempted for L-mode discharges using a $\chi^2$ minimization via the Levenberg-Marquardt Algorithm \cite{Howard2012QuantitativePlasma}. This approach suffers from the risk of finding \emph{local} $\chi^2$ minima rather than \emph{global} ones, depending on initial guesses of model parameters. It also lacks of rigorous uncertainty quantification and model selection, as discussed by Chilenski \emph{et al.}~\cite{Chilenski2019OnProfiles}, who adopted Bayesian techniques with synthetic data to demonstrate the critical importance of avoiding under- or over-fitting\cite{Chilenski2019OnProfiles}. 
Unfortunately, synthetic data often makes it difficult or impossible to explore issues of \emph{model inadequacy} that may result from over-simplified experimental analysis, inaccurate atomic rates, discrepancies of detector models and other important factors\cite{Morrison2018RepresentingApproach}. For example, the omission of sawtooth modeling or the time-dependence of atomic rates, as in previous C-Mod work\cite{Howard2012QuantitativePlasma,Chilenski2019OnProfiles,Chilenski2018EfficientExperiments}, may prevent the optimal model complexity to emerge. 

In this work, we demonstrate parameter estimation and model selection using a forward model that has been improved in both its physics fidelity and its iterative capability. We describe our approach on experimental data, thus parting from any idealizations of synthetic datasets. Our fully-Bayesian framework employs nested sampling algorithms and introduces faster forward modeling, higher physics fidelity, improved treatment of experimental measurements and nuisance parameters, as well as more rigorous combination of diagnostic instruments. We attempt to model an extensive range of experimental details to improve comparison of experiment with neoclassical, gyro-fluid and gyrokinetic models.

The structure of this paper is as follows. In section~\ref{sec:setup}, we describe the typical experimental setup for impurity transport experiments at Alcator C-Mod and we introduce a high-performance I-mode discharge that will be used to demonstrate our inference methods. A detailed description of our inference framework is given in section~\ref{sec:methodology}. In section~\ref{sec:results} we compare results of experimental inferences to neoclassical and turbulence modeling. The latter is analyzed in more detail in section~\ref{sec:modeling}. Finally, in section~\ref{sec:discussion} we discuss and summarize this work. Additional details on numerical schemes and experimental data are provided in the appendices.

%
%
\section{Experimental Setup and Measurements} \label{sec:setup}
The analysis methods presented in the next section are applicable to a wide range of discharges and we therefore describe the general experimental setup that can be used for impurity transport inferences. We focus on data from Alcator C-Mod\cite{Hutchinson1994FirstAlcator-C-MOD}, a compact ($a=0.22$ m, $R = 0.68$ m), high field, diverted tokamak that operated up to the end of 2016 with up to 5.5 MW of ion cyclotron resonance heating (ICRH). C-Mod parameters have ranged within $B_T=[2.0-8.1]$ T, $I_p=[0.4, 2.0]$\,$\textrm{MA}$, $n_e(r=0)=[0.2- 6.0]\times10^{20}$\,$\textrm{m}^{-3}$ and $T_e(0)$ up to $9$ keV\cite{Greenwald201420Tokamak}. 

On Alcator C-Mod, a multi-pulse LBO system has been used for transient impurity studies for several years \cite{Howard2011CharacterizationSystem, Howard2012QuantitativePlasma, Howard2014ImpurityPlasmas, Rice2014X-rayPlasmas, Rice2015X-rayPlasmas}. Here, we consider experiments where a CaF$_2$ film was ablated and calcium ($Z_{Ca}=20$) transport was studied by diagnosing impurity emission with a combination of spectroscopic measurements. A significant advantage of Ca over other ions is that it is non-intrinsic and non-recycling, and it is also simple to ablate from a LBO slide\cite{Howard2011CharacterizationSystem}.


An initial condition for particle transport reconstructions was provided by an optical fiber viewing the edge of the plasma, near the injection location of the LBO. A bandpass filter limited measured brightness to the $420 \pm 10$ nm range, providing a proxy measurement of Ca-I emission\cite{Howard2012QuantitativePlasma}. This system offers high signal-to-noise ratio (SNR) and sometimes provides evidence of CaF$_2$ clusters entering the plasma; this often correlates with irregularities in signals on several other spectrometers. 

For our inferences, a central role is played by C-Mod's XICS diagnostic\cite{Reinke2012X-rayResearch}, which has 3 ms readout time and typical signal integration over 3 ms intervals, for a total time resolution of 6 ms. The two crystals of this system are normally configured to observe lines from the H- and He-like charge states of argon. Doppler broadening and Doppler shift of the emission spectra provide radial profiles of ion temperature ($T_i$) and toroidal rotation ($v_{\phi}$) following tomographic inversion\cite{Reinke2012X-rayResearch,Rice2013Non-localPlasmas}. 

For impurity transport studies, one of the two XICS crystals could be substituted with one capable of viewing the Ca$^{18+}$ (He-like) spectrum, often discretizing the detector spatial coverage into 32 independent spatial chords. 
In standard C-Mod discharges, the brightest Ca line in the core is the 1s$^2$ $^1S_0$ - 1s2p $^1P_1$ resonance (w) line at 3177.26 $m$\AA, and past inferences of impurity transport coefficients in Refs.\cite{Howard2012QuantitativePlasma,Howard2014ImpurityPlasmas,Chilenski2018EfficientExperiments} relied mostly on its measurement using existing spectral analysis routines\cite{Reinke2013OperationC-Mod}. As described in section~\ref{sec:spectroscopy_methods}, fitting and uncertainty quantification for the z line (1s$^2$ $^1S_0$ - 1s2p $^1P_1$, 3211.13 $m$\AA) for impurity transport studies has also recently become routine.

Finally, a single-chord VUV spectrometer (XEUS) focused on the 10-70\,\AA\,range can be used to view multiple emission lines for Li-like Ca\cite{Reinke2010VacuumTokamak}. In recent years, a companion Long-wavelength Extreme Ultraviolet Spectrometer (LoWEUS) spectrometer\cite{Lepson2016ResponsivitySpectrometer} has allowed emission in the 100-300$\,$\AA$\,$region to also be analyzed. 
Unfortunately, LoWEUS data are available only in a small subset of C-Mod data. While the spectral range of these spectrometers typically offers valuable observations of Ca$^{16+}$ and Ca$^{17+}$ at $r/a>0.7$, their relatively low time resolution (2 ms) does not usually allow clear observation of the signal rise phase following LBO injections.  

In this paper, we focus on a sawtoothing I-mode discharge run with $B_t \approx 5.5$ $T$ and $I_p \approx 1.0$ $MA$, reaching $T_e\approx 5$ keV and $n_e \approx 1.7 \times 10^{20}$ $m^{-3}$ on axis. Using the neoclassical prediction of Sauter \emph{et al.}\cite{Sauter2002Erratum:2834}, we estimate $Z_{eff}\approx2.1 \pm 0.2$ during the current flat-top and assume radial $Z_{eff}$ profiles to be flat, consistently with past Visible Bremsstrahlung measurements on C-Mod\cite{Fiore2006InternalC-Mod}. This leads to a volume-averaged (dimensionless) collisionality, defined as $\nu_{ei}=0.1$ $Z_{eff}$ $\langle n_e \rangle $ $R/\langle T_e \rangle ^2 $\cite{Angioni2007ScalingObservations}, of $0.9 \pm 0.2$, a relatively low value for standard Alcator C-Mod operation. This discharge presents a temperature pedestal, but the electron density decreases smoothly across the LCFS. This is consistent with the typical phenomenology of I-mode, which is characterized by good energy confinement, no Edge-Localized Modes (ELMs), a weakly-coherent $\sim 100-300$ kHz mode near the LCFS, high levels of particle transport and low impurity retention\cite{Whyte2010I-mode:C-Mod, Rice2015CorePlasmas}. 

\begin{figure}[ht]
		\centering
	\begin{subfigure}{0.4\textwidth}
		\centering
		\includegraphics[width=0.99\linewidth]{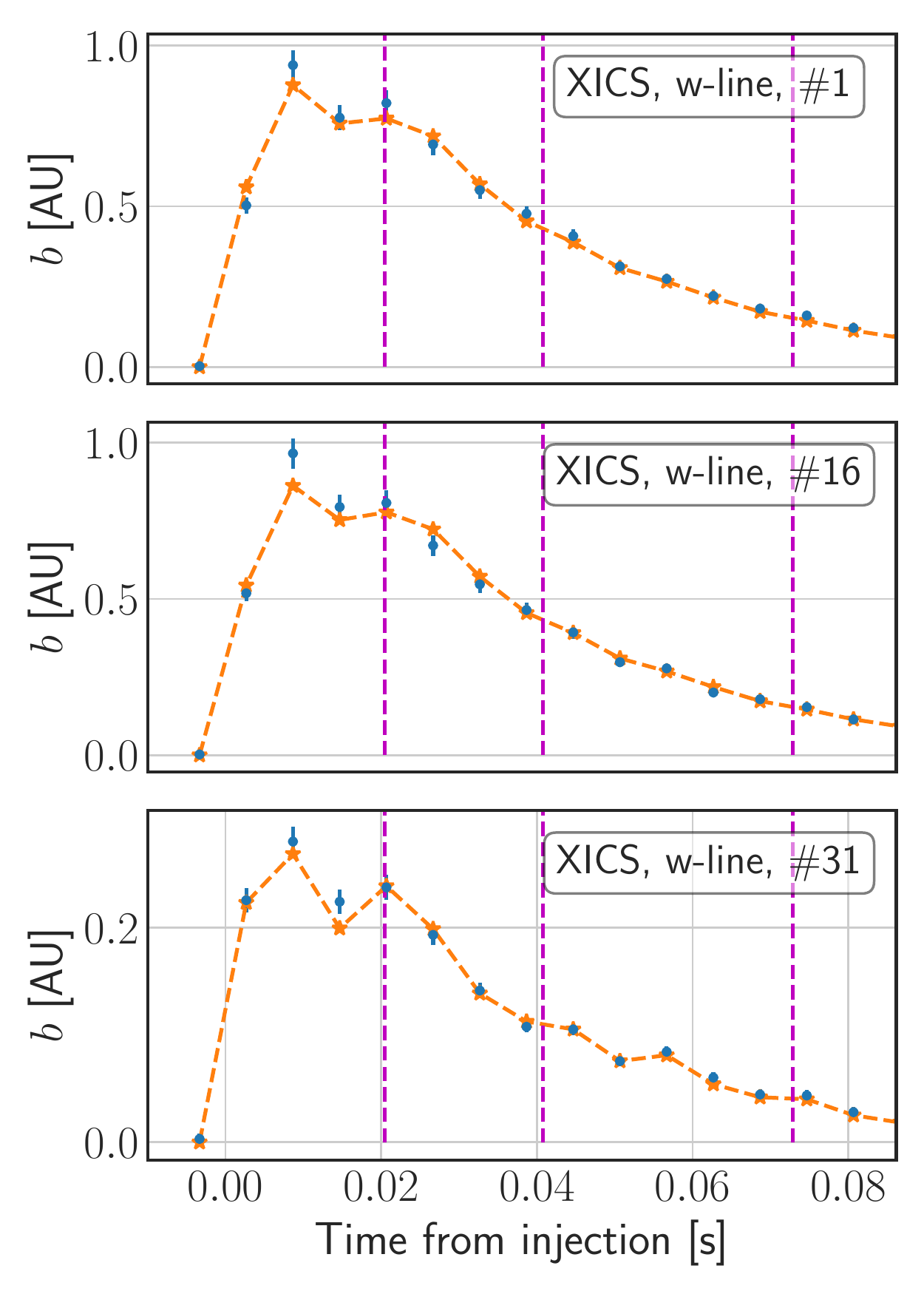}
		\caption{ }
	\end{subfigure}
	\begin{subfigure}{0.4\textwidth}
		\centering
		\includegraphics[width=0.99\linewidth]{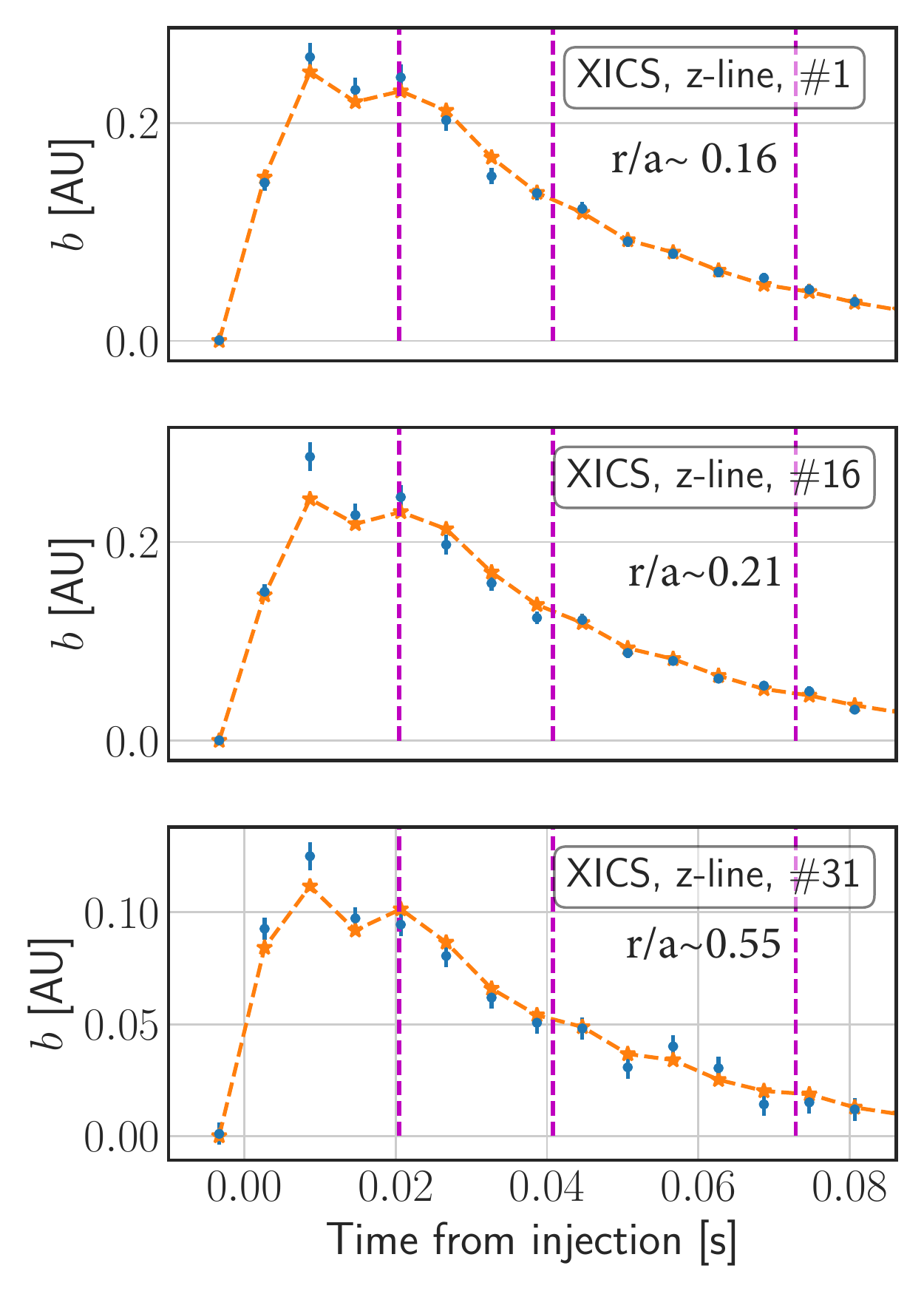}
		\caption{}
	\end{subfigure}
	\caption{XICS Ca$^{18+}$ w (left) and z line (right) signals from sample chords following a LBO injection into the I-mode discharge of interest. Tangency radii are indicated within plots on the right. Orange dashed lines show fits from the impurity transport inference described in section~\ref{sec:results}. Vertical magenta dashed lines indicate times at which sawteeth occur.}
	\label{fig:signals_decays}
\end{figure}

Fig.~\ref{fig:signals_decays} shows w and z line brightness signals from sample XICS chords following a Ca LBO injection during the current flat-top of this discharge. Orange lines in Fig.~\ref{fig:signals_decays} show signal fits from the impurity transport inference described in section~\ref{sec:results}. Tangency radii for these chords are shown in the plots on the right hand side and are below the midplane for low-numbered chord and above it for high-numbered ones. In high performance discharges such as the presented I-mode, He-like Ca emission is collected via XICS up to $r/a\approx 0.8$, outside of which XEUS still measures several Li-like Ca lines.

\begin{figure}[ht]
	\centering
	\includegraphics[width=0.5\textwidth]{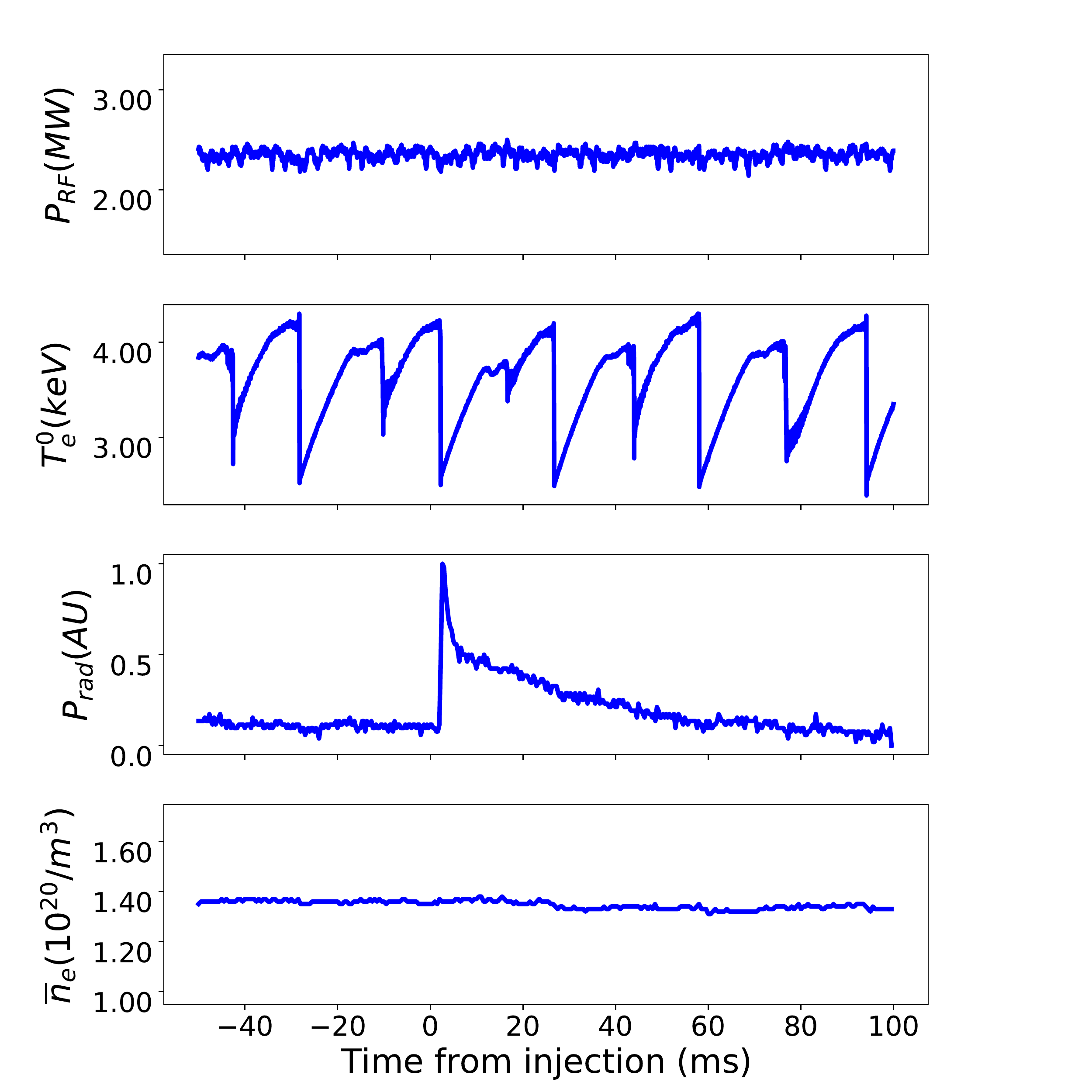} 
	\caption{Typical time series of radio-frequency power ($P_{RF}$), near-axis electron temperature ($T_e^0$), radiated power ($P_{rad}$) and line-averaged electron density ($n_e$) for the I-mode discharge described in the main text. The LBO injection is seen not to perturb the electron density.}
	\label{fig:shots_time_series}
\end{figure}

In Fig.~\ref{fig:shots_time_series}, we show time traces of H-minority Ion Cyclotron Resonance Heating (ICRH) power ($P_{RF}$) with a near-axis resonance, central electron temperature ($T_{e,0}$), total radiated power from bolometry ($P_{rad}$) and line-averaged density ($\bar{n}_e$) from the same I-mode discharge. As it is often the case in C-Mod, sawteeth modulate core temperatures and they are known to also affect impurity profiles. Previous work on C-Mod focused on inferring ``sawtooth-averaged'' impurity transport \cite{Howard2012QuantitativePlasma,Chilenski2017ExperimentalSimulations}, which may be a sufficiently good approximation to background transport coefficients in low-performance discharges with small sawteeth. In order to examine higher-$T_e$ discharges, broadening the range of analyzed confinement states and enabling higher SNR from spectroscopic signals in the plasma edge, some form of sawtooth modeling is necessary, and will be the described in section~\ref{sec:imp_transport}. 

A requirement for our transport inferences to be meaningful for validation purposes is that the plasma is quasi-steady and mostly unperturbed by LBO injections. This is illustrated in Fig.~\ref{fig:shots_time_series}, where the line averaged density (lowest panel, measured by a Two-Color Interferometer) appears unaffected by the CaF$_2$ influx (shown by radiated power, measured via bolometry).

\begin{figure}[ht]
		\centering
	\begin{subfigure}{0.25\textwidth}
		\centering
		\includegraphics[width=0.99\linewidth]{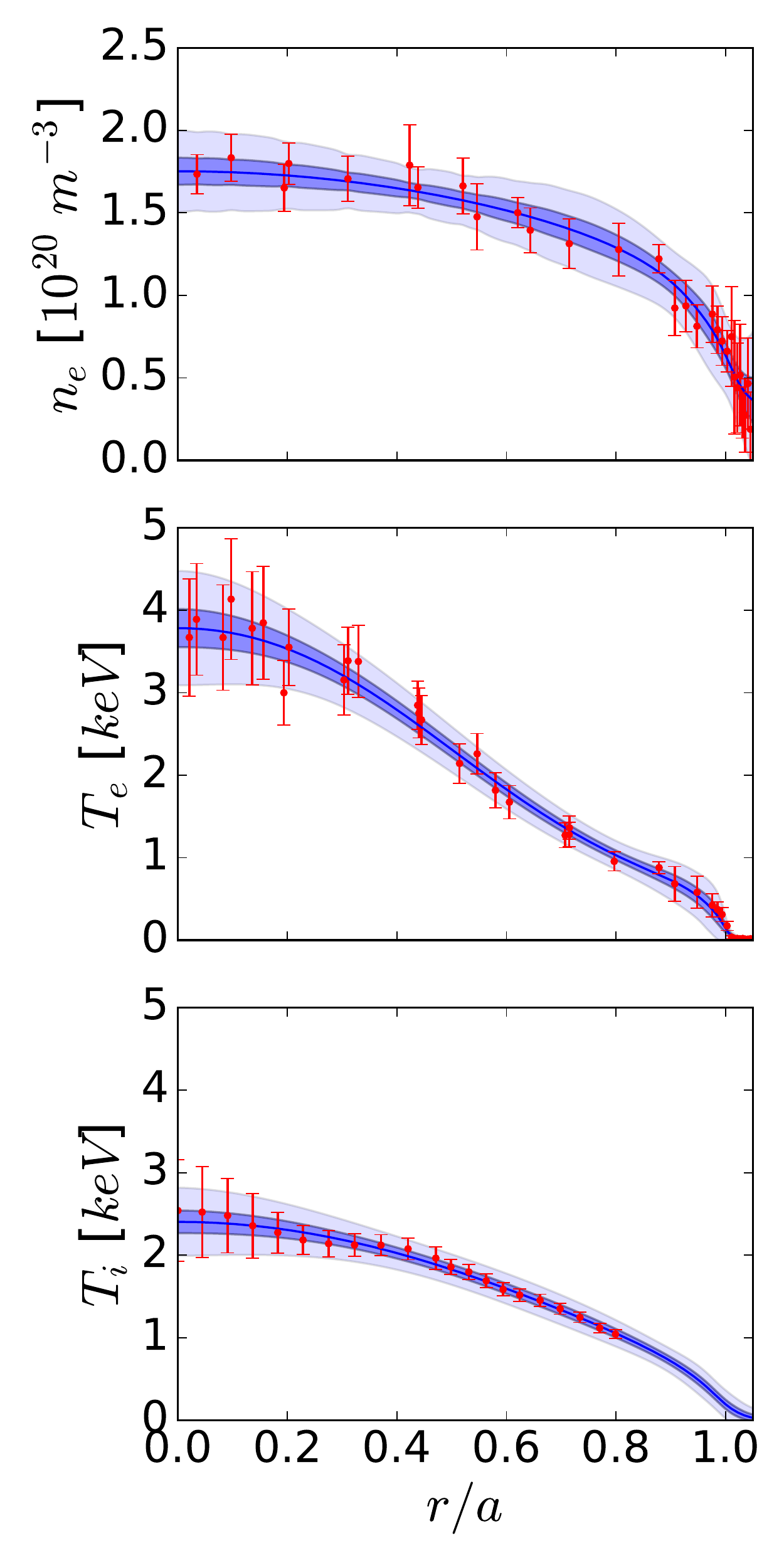}
		\caption{ }
	\end{subfigure}
	\begin{subfigure}{0.25\textwidth}
		\centering
		\includegraphics[width=0.99\linewidth]{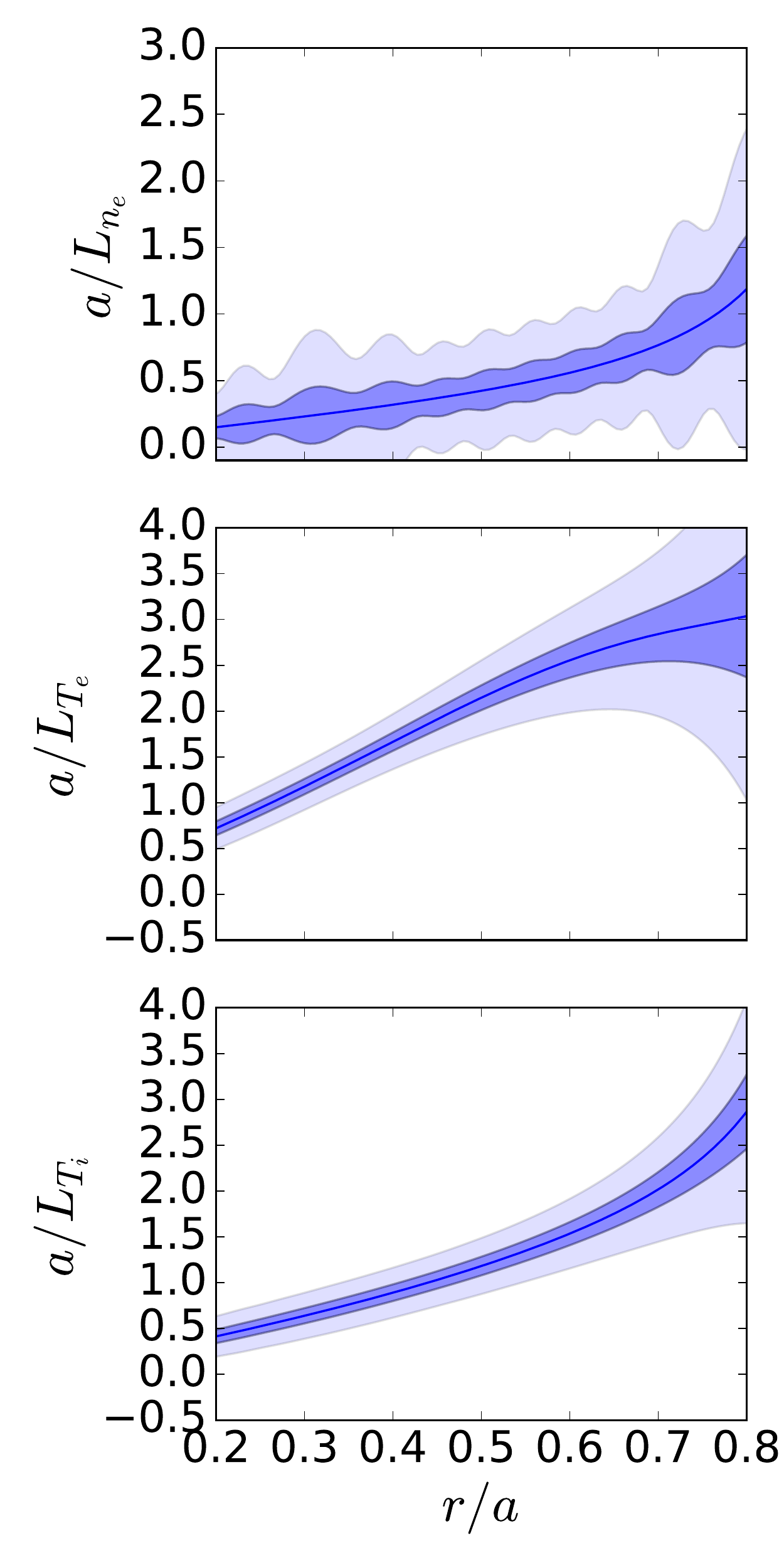}
		\caption{}
	\end{subfigure}
	\caption{Time-averaged kinetic profiles for the I-mode discharge, fitted via Gaussian Process Regression. In (b), we focus on gradient scale lengths between $r/a=0.2$ and $r/a=0.8$.}
	\label{fig:MITIM1_paper_kinetic_profiles}
\end{figure}

We adopt different kinetic profile fitting methods for impurity transport forward modeling and for theoretical (neoclassical, gyro-fluid and gyrokinetic) predictions. For forward modeling, we make use of a parametric time-dependent Radial Basis Functions (RBF), taking care to avoid smoothing of sawteeth and to enforce pedestal structure. Use of fast Electron Cyclotron Emission data allows us to fit very well the $T_e$ modulation, while for $n_e$ we rely on Thomson Scattering data at lower time resolution.

For theoretical modeling, we adopt Gaussian Process Regression (GPR), which minimizes user bias by inferring hyperparameters corresponding to fits of highest probability\cite{Chilenski2015ImprovedRegression}. In Fig.~\ref{fig:MITIM1_paper_kinetic_profiles}, we show time-averaged kinetic profile fits ($n_e$, $T_e$ and $T_i$) and corresponding normalized inverse gradient scale lengths ($a/L_{n_e}$, $a/L_{T_e}$ and $a/L_{T_i}$, with $L_X=-X/(dX/dr)$) that result from this procedure using data from the time interval $[1.15,1.3]$s. In the absence of experimental $T_i$ measurements in the pedestal, we assume high collisional coupling between electrons and ions and take $T_i\approx T_e$ in this region, as experimentally observed in past work\cite{Churchill2015PoloidalBarriers}. We quantify GPR uncertainties by propagating diagnostic uncertainties and data scattering in time via the Law of Total Variance and MCMC sampling\cite{Bishop2006PatternLearning}, using a Gibbs kernel with $\tanh$ length scale, as suggested in Ref.\cite{Chilenski2015ImprovedRegression}.

%
%
\section{Methods} \label{sec:methodology}
\begin{figure}[ht]
	\centering
	\includegraphics[width=0.99\textwidth]{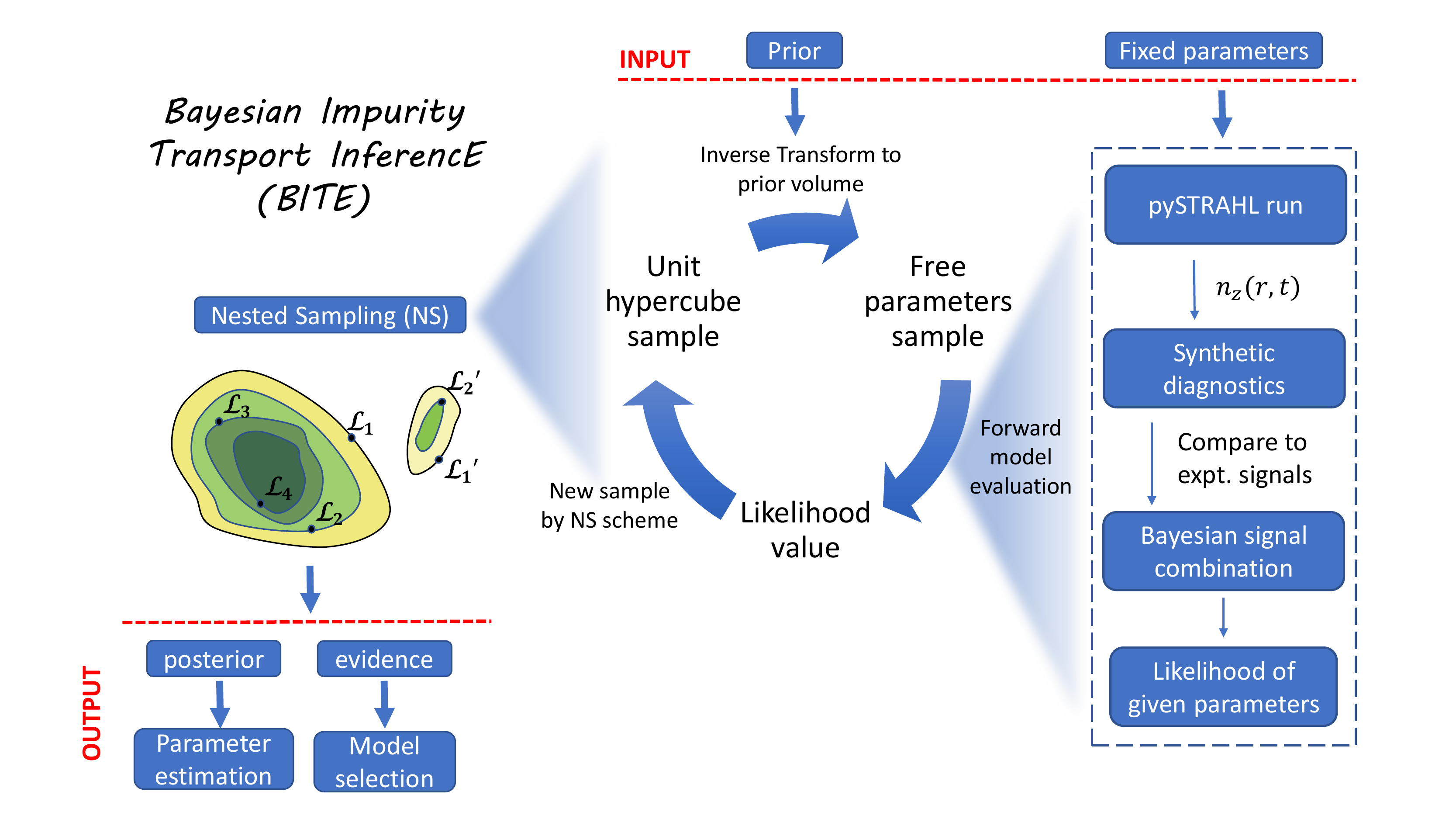}
	\caption{Overview of the components of the \texttt{BITE} framework to infer radial profiles of impurity transport coefficients.}
	\label{fig:methods_diagram}
\end{figure}

In this section, we describe the Bayesian Impurity Transport InferencE (\texttt{BITE}) framework developed in this work and broadly represented in Fig.~\ref{fig:methods_diagram}. At the highest level, \texttt{BITE} uses the nested sampling algorithm to iteratively test predictions from possible parameter choices. In doing so, it repeatedly runs a forward model for impurity transport, represented in the right hand side box of Fig.~\ref{fig:methods_diagram}. In what follows, we describe in greater depth each aspect of this diagram.

\subsection{Impurity Transport Forward Modeling} \label{sec:imp_transport}

To forward-model the radial transport of an impurity ``I'', we make use of the continuity equations for each impurity charge state  ``Z'',
\begin{equation} \label{eq:nz_cont}
\frac{\partial n_{I,Z}}{\partial t} = - \frac{1}{r} \frac{\partial}{\partial r} \left( r \vec{\Gamma}_{I,Z} \right)+Q_{I,Z},
\end{equation} 
with the radial particle flux $\Gamma_{I,Z}$ being
\begin{equation} \label{eq:rad_flux} 
\Gamma_{I,Z}=-D \frac{\partial n_{I,Z}}{\partial r}+v \ n_{I,Z}.
\end{equation}
The source/sink term $Q_{I,Z}$ couples the transport equation of each ionization stage with the neighboring ones via ionization and recombination processes, for which we use rates from the Atomic Database Analysis System (ADAS)\cite{Summers2011AtomicProject}\footnote{We take thermal charge exchange processes to be negligible, assuming that neutrals do not penetrate much further than the LCFS in the high-density C-Mod conditions.}. We consider flux-averaged transport coefficients and label flux surfaces based on the volume they enclose, using $r = \sqrt{V/(2\pi^2 R_{axis})}$. 

Theoretical transport models can often be used to predict coefficients $D_0$ and $v_0$ with the respect to the low-field-side (LFS) density $n_0$. To compare these results, on a minor radius coordinate $r_m$, to those from forward modeling, on the $r$ coordinate defined above, one can perform the following transformations\cite{Angioni2014TungstenModeling}:
\begin{equation} \label{eq:corrections}
D = D_0 \frac{n_0}{\langle n\rangle} \left(\frac{\partial r}{\partial r_m}\right)^2  \qquad \textrm{and} \qquad v = v_0 \frac{n_0}{\langle n\rangle} \frac{\partial r}{\partial r_m} + D \frac{\partial }{\partial r_m} \left( \ln \frac{n_0}{\langle n\rangle}\right)
\end{equation}
where factors of $n_0/\langle n\rangle$ account for poloidal density asymmetries, expected to be significant for heavy ions\cite{Reinke2012PoloidalC-Mod,Odstrcil2018TheUpgrade}. The $n_0/\langle n\rangle$ ratio may be estimated using a neoclassical code or, more simply, an analytical prediction of centrifugal effects\cite{Odstrcil2018TheUpgrade}. 
A detailed description of this correction procedure, which can modify $D$ and $v$ estimates by up to 50\% for Ca ions in C-Mod plasmas, can be found in Appendix A of Ref. \cite{Angioni2014TungstenModeling}. 

Eqs.~\ref{eq:nz_cont} and~\ref{eq:rad_flux} represent a system of coupled differential equations for all charge states. We solve this using a new \texttt{pySTRAHL} package, which builds on the well-known \texttt{STRAHL} code \cite{Dux2006STRAHLManual}, with which it has been thoroughly benchmarked. \texttt{pySTRAHL} is written in Fortran 90 and couples to Python via the \texttt{f2py} package, maintaining modeling capabilities of the original \texttt{STRAHL} while avoiding the redundant input/output (I/O) normally required to iterate over \texttt{STRAHL} inputs. At present, \texttt{pySTRAHL} iterations can either be initialized directly in Python or by reading the output of a first \texttt{STRAHL} run which is written to a NETCDF file and can later be read into computer memory. Following this initialization stage, \texttt{pySTRAHL} iterations only require modification of input variables in Python.

With either initialization method, \texttt{pySTRAHL} requires similar inputs to \texttt{STRAHL}: plasma geometry, atomic rates, kinetic profiles, impurity sources, recycling and scrape-off-layer (SOL) parameters and transport coefficients (diffusion and convection as a function of radius). The latter are the quantities that we wish to infer, and are therefore iterated over according to a chosen algorithm. At all iterations, the values of $D$ and $v$ at a specified number of knots are interpolated using Piecewise Cubic Hermite Interpolating Polynomial (PCHIP) splines\cite{Fritsch1980MonotoneInterpolation}, which enforce monotonicity between knots and avoid over-shooting. In \texttt{BITE}, these profiles are typically summed with an additional Gaussian feature in the $v$ edge profile, representing the expected particle pinch in the pedestal region. This Gaussian is typically centered at the LCFS and its width is allowed to be a free parameter. This parameterization enables one to represent sharp changes in the $v$ profile without necessitating a large number of additional free spline knots or detailed experimental data in the pedestal. 

In order to compare transport models for impurity propagation through the pedestal, we allow $v/D$ to scale with the $Z$ value of each charge state outside of a specified radial region. Such scaling is expected by neoclassical theory\cite{Pedersen2000RadialC-Mod,Helander1998NeoclassicalPlasma}\footnote{A weaker neoclassical dependence of $D$ on $Z$ is also expected, but we did not attempt to model this.} and has been observed in previous experimental work\cite{Putterich2011ELMUpgrade}. The radial location outside of which $v$ is taken to scale with $Z$ is determined by a parameter that we normally fix for simplicity to be $r/a=0.9$ (approximately the pedestal top). One can then test how results compare to those obtained by standard runs with no $v$ charge-state dependence. 

In sawtoothing discharges, the application of a sawtooth model can be important. \texttt{STRAHL} allows one to flatten all charge state density profiles inside of the sawtooth mixing radius at given crash times; this phenomenological picture was suggested by experimental observations by Seguin \emph{et al.}\cite{Seguin1983EffectsTokamaks}. A drawback of this simple sawtooth model is that it creates sharp density gradients at each sawtooth crash, thus demanding high temporal resolution to remain numerically stable. By implementing a smoother (neither more nor less physical) sawtooth crash model, such requirements have been alleviated, effectively halving the runtime of \texttt{pySTRAHL} simulations. Crashes were implemented using complementary error functions as  
\begin{equation}
    n_z(r) \leftarrow \frac{\bar{n}_z}{2} \erfc{\left(\frac{r- r_{mix}}{w}\right)} + \frac{n_z(r)}{2} \erfc{\left(\frac{r_{mix} - r}{w}\right)}
\end{equation}
where the left-pointing arrow shows how $n_z(r)$ changes across the crash. Here, $n_z$ is the charge state density, $\bar{n}_z$ is its average evaluated between the magnetic axis and the mixing radius before sawtooth crashes and $w\approx 1$ cm is a ``crash region width'', which sets the spatial smoothness of the sawtooth crash.

The output of \texttt{pySTRAHL} consists of time-dependent charge state densities for all ionization stages of an impurity ion in each of the particle reservoirs. By summing densities in the plasma, wall and divertor regions, we have ensured that particle conservation is maintained in our simulations. At all iterations, charge state distributions are combined with atomic rates to compute local emissivities. These are then line-integrated to compare to spectroscopic diagnostic brightness. Radial profiles of transport coefficients are then modified to better match experimental data (based on algorithmic schemes) and are given as inputs to \texttt{pySTRAHL}, allowing for a new comparison with experimental signals. 

\texttt{pySTRAHL} offers major advantages over standard \texttt{STRAHL} operation for iterative work. First of all, by caching spatio-temporal grids, atomic rates, geometry and other fixed parameters at an initialization stage, we obtain a speed improvement of approximately $70\%$, bringing our runtime down to $\approx 150$ ms per simulation for the I-mode case discussed in this work\footnote{These times refer to cluster nodes using Intel(R) Xeon(R) CPU E5-2683 v4, 2.1 GHz.}. The speed-up can be even more significant for runs where input/output (I/O) operations would normally take a larger fraction of the \texttt{STRAHL} runtime. Moreover, by avoiding I/O at every iteration we are able to efficiently parallelize our inference and make use of hundreds of CPUs. Without this capability, reaching convergence with the advanced algorithms described in section~\ref{sec:bayesian}, requiring tens of millions of iterations at times, would be unfeasible.

\subsection{Spectroscopic analysis} \label{sec:spectroscopy_methods}
\begin{figure}[ht]
	\begin{subfigure}{0.5\textwidth}
		\centering
		\includegraphics[width=0.99\linewidth]{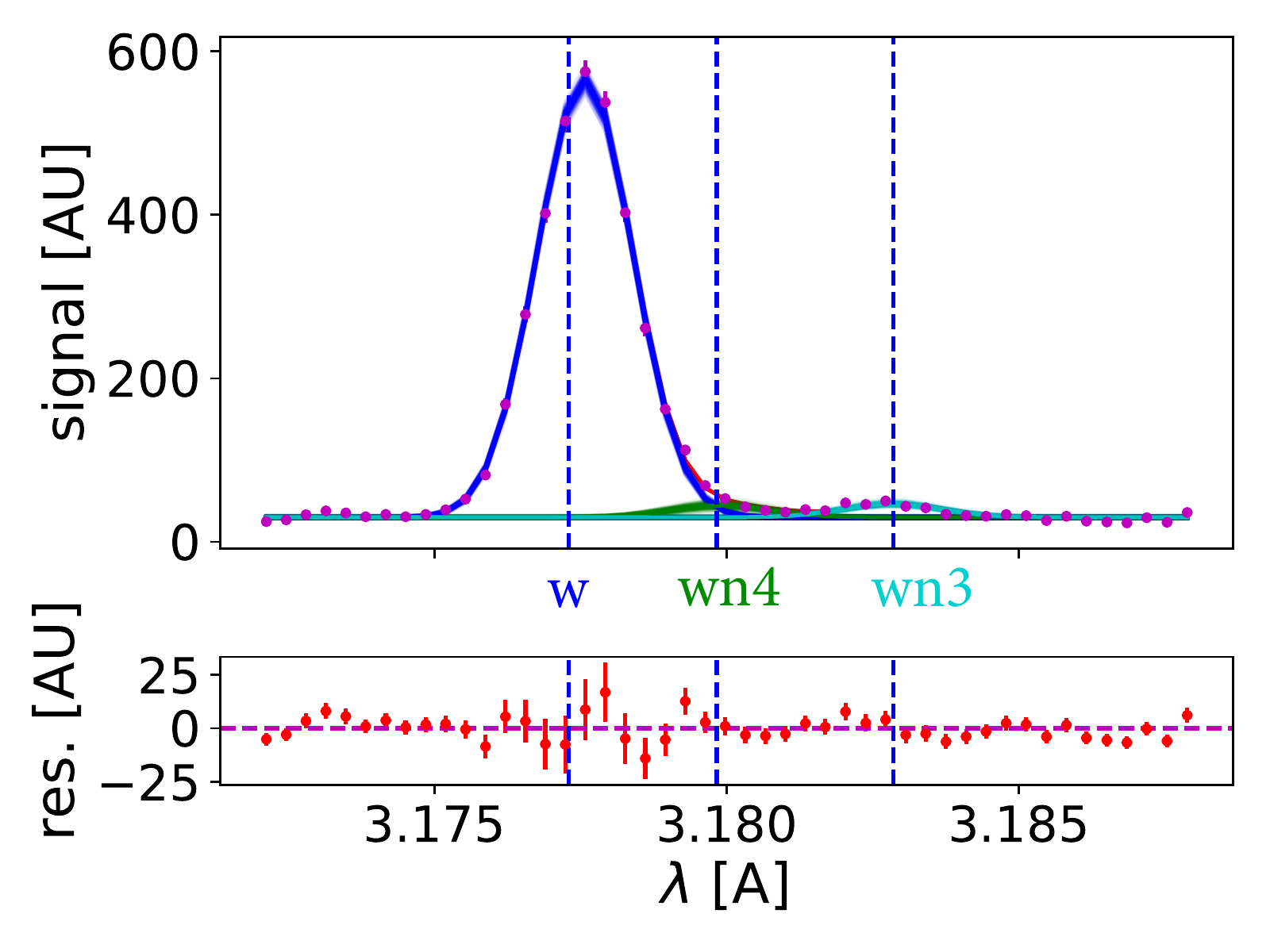}
		\caption{ }
		\label{fig:bsfc_w}
	\end{subfigure}
	\begin{subfigure}{0.5\textwidth}
		\centering
		\includegraphics[width=0.99\linewidth]{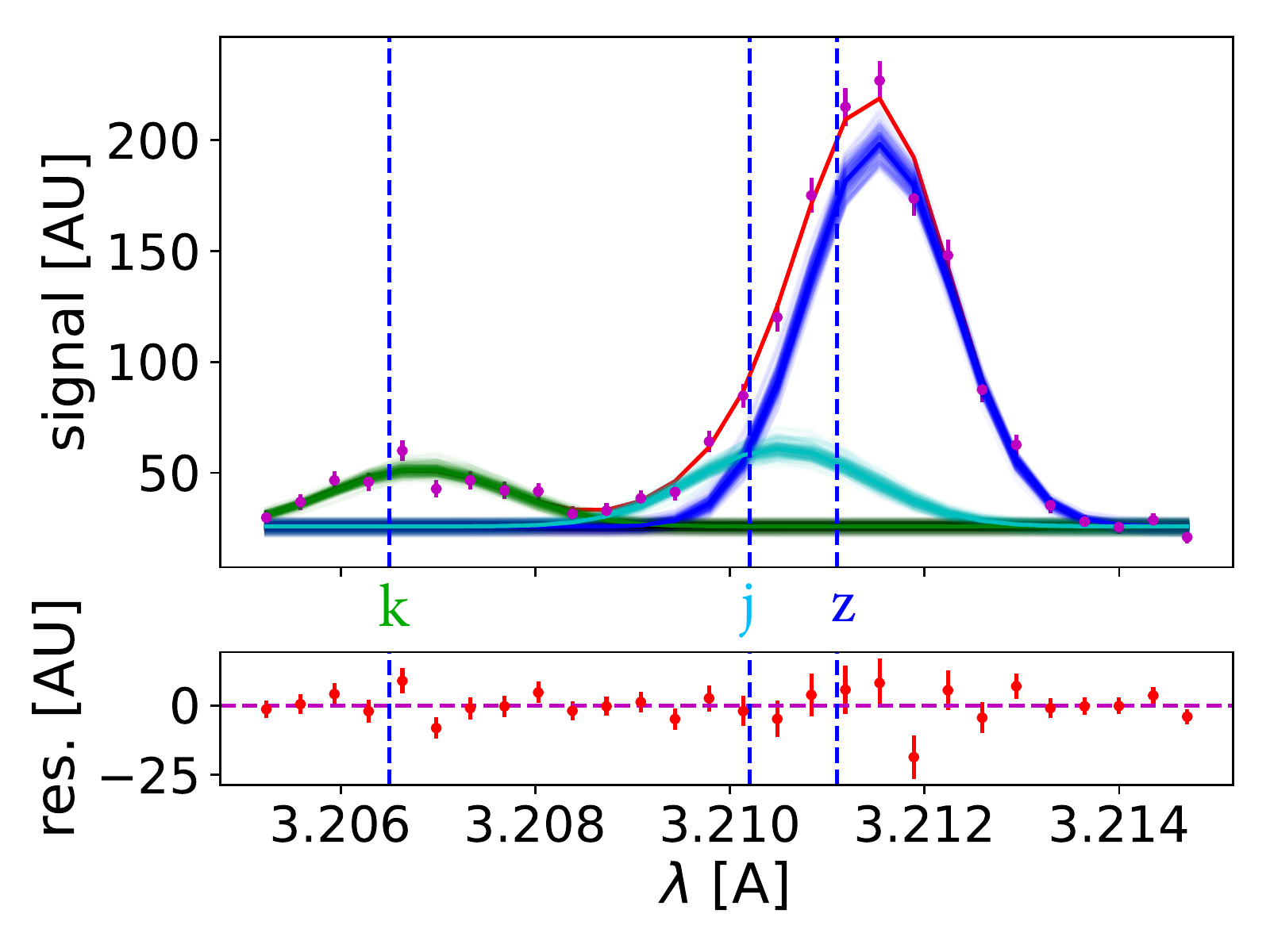}
		\caption{}
		\label{fig:bsfc_z}
	\end{subfigure}
	\caption{Typical fits of the XICS He-like Ca spectra. In (a), the brightest line is the w resonance; in (b), it is the z forbidden line. These fits correspond to $\sim1\,\text{ms}$ after the LBO injection in the I-mode plasma discussed in the main text. Bottom panels show spectral fitting residuals. }
	\label{fig:bsfc_fits}
\end{figure}

The accurate analysis and interpretation of atomic spectra, particularly for high-resolution data including overlapping lines, is paramount for transport inferences. The development of the Bayesian Spectral Fitting Code (BSFC) has recently offered new opportunities to analyze XICS data on C-Mod where traditional spectral fitting tools would find it difficult to effectively subtract overlapping satellite lines from a primary line of interest \cite{Cao2019BayesianQuantification}. BSFC addresses this problem with a decomposition of each lineshape into a Hermite polynomial series and using Bayesian sampling techniques, analogous to those used in \texttt{BITE}. This allows robust and rigorous truncation, as well as uncertainty quantification. In Fig.~\ref{fig:bsfc_fits} we show spectral fits obtained for the I-mode discharge of interest at $t\approx 1.21$ s, following the Ca injection at $t=1.2$ s. Fig.~\ref{fig:bsfc_w} shows $n\geq4$ satellite lines blending with the 1s$^2$-1s2p resonance w line (rest wavelength of $3177.3$ m\AA). In Fig.~\ref{fig:bsfc_z}, we show the forbidden z line (rest wavelength of $3211.1$ m\AA) blending with the j line, whose amplitude, width and Doppler shift are physically constrained by the nearby k line. The spectra in Fig.~\ref{fig:bsfc_w} and~\ref{fig:bsfc_z} are measured on two adjacent detector submodules, whose accurate cross-calibration has been demonstrated in previous work (e.g. Ref.~\cite{Rice2014X-rayPlasmas}) where w and z line spectra could be simultaneously matched by atomic modeling. Therefore, we can make use of the relative amplitude of these lines without running into cross-calibration issues. Since the diagnostic is not absolutely calibrated, we normalize both the w and z line signals to the maximum signal observed across all spatial chords for the two lines together. 

Measuring both the w and z line emission offers a powerful way to constrain impurity transport, since these lines arise from different physical processes: the upper level of the w transition is populated via collisional excitation of the ground state of the He-like ion; for the z transition, the upper state is instead populated via radiative recombination of the H-like ion and inner shell ionization of Li-like one. As one might expect, these processes have different $T_e$ dependence, with z line emissivity becoming stronger at lower $T_e$ and signals extending to the pedestal\cite{Rice2007ImpurityPlasmas,Rice2013Non-localPlasmas}. In order to compute atomic rates, we make use of formulae and coefficients provided by Mewe and Schrijver \cite{MeweR.andSchrijer1978Helium-likeIntensities}, which have previously been shown to match C-Mod XICS spectra with great accuracy\cite{Rice2015X-rayPlasmas}. This calculation requires detailed knowledge of the H-, He- and Li-like ion populations, which are output by \texttt{pySTRAHL} after each run; therefore, w and z line rates are computed at every iteration in \texttt{BITE}. The ratio of w and z line intensities effectively constrains whether the impurity of interest is recombining, ionizing or is in steady-state at the local temperature\cite{Gabriel1972DielectronicLines}, providing a strong constraint on plasma transport properties. Optimizing the calculation of these rates in Fortran 90, we have been able to reduce the runtime to $75$ ms, making our entire forward model execution (including a \texttt{pySTRAHL} run and the application of synthetic diagnostics) shorter than $300$ ms.

For VUV spectrometers, the lower spectral resolution does not permit advanced fitting procedures and we resort to standard binning of known atomic lines. We combine the brightness of adjacent lines for which ADAS photon emissivity coefficients are available, so as to increase the signal-to-noise ratio. For CaF$_2$ LBO injections, we found the ($1.8683$,$1.8727$) nm, ($1.9775$, $1.9632$, $1.9790$) nm and ($2.0122$, $2.0289$) nm Li-like Ca groupings to be clearly visible on the XEUS spectrometer\cite{Reinke2010VacuumTokamak}.

To account for possible systematic errors, we apply a minimum relative uncertainty of $\pm5\%$ to all brightness values as in past analysis\cite{Howard2012QuantitativePlasma,Chilenski2017ExperimentalSimulations}. For VUV signals, uncertainties are most effectively estimated from signal scatter before LBO injections.

\subsection{Bayesian Inference} \label{sec:bayesian}
Bayesian analysis methods are commonly applied in several branches of Physics and Engineering, but are relatively less known in fusion research. For this reason, we introduce some key Bayesian concepts in this section. 

Consider a framework to infer parameters $\theta$ (vector symbols are omitted in the following) given experimental data $\mathcal{D}$ and a model $\mathcal{M}_i$. Bayes' formula then reads
\begin{equation} \label{eq:Bayes}
p(\theta, \gamma | \mathcal{D}, \mathcal{M}_i) = \frac{ p(\mathcal{D}|\theta, \gamma, \mathcal{M}_i) p(\theta, \gamma| \mathcal{M}_i)}{\mathcal{Z}(\mathcal{D}|\mathcal{M}_i)}, 
\end{equation}
where 
\begin{equation} \label{eq:evidence}
\mathcal{Z} (\mathcal{D}|\mathcal{M}_i) = \int p(\theta, \gamma| \mathcal{M}_i) p(\mathcal{D}|\theta, \gamma, \mathcal{M}_i) d\theta d\gamma
\end{equation}
is the \emph{evidence} (or \emph{marginal likelihood}) of model $\mathcal{M}_i$. Here, $\theta$ is a vector containing the parameters that we wish to infer and $\gamma$ is another vector containing nuisance parameters, i.e. parameters that are needed by our model evaluation, but whose values are not of physical interest (see section~\ref{sec:sig_modeling}). Any ``choice'' of parameters $\{\theta,\nu\}$ may be referred to as a parameter \emph{sample}. The term $p(\theta, \gamma| \mathcal{M}_i)\equiv p(\theta, \gamma)$ (a probability density for the parameters, given a specific model) is generally referred to as the prior (see section~\ref{sec:priors}), while $p(\mathcal{D}|\theta, \gamma, \mathcal{M}_i)\equiv \mathcal{L}(\mathcal{D}|\theta)$ is the likelihood. The left hand side, $p(\theta, \gamma | \mathcal{D}, \mathcal{M}_i)\equiv p(\theta|\mathcal{D})$, is the posterior. Note that in order to find the parameters $\theta$, $\gamma$ that maximize the posterior density function (the ``Maximum A-Posteriori'', or MAP, estimate), we can ignore $\mathcal{Z}$, since this is not a function of these parameters. However, while the integration required to compute $\mathcal{Z}$ tends to be a computationally expensive task, its value (or approximations thereof) is the fundamental \emph{quantitative} Bayesian metric to compare how much the data support one mathematical model versus another. 
The comparison of $\mathcal{Z}$ values (or their equivalent normalization) is referred to as \emph{model selection}. A complementary task to model selection is \emph{parameter estimation}, which attempts to find the best parameters for a given model. Both model selection and parameter estimation are important for \emph{model validation}, although the former is often not treated in detail due to its inherent complexity. 

In the absence of good estimates of the Bayesian evidence, one may resort to some of its approximations, e.g. the Akaike Information Criterion (AIC) or the Bayesian Information Criterion (BIC), which however are derived by making questionable assumptions on the form of the (unknown) posterior distribution\cite{Trotta2008BayesCosmology}. Note that the $\chi^2$ is a metric describing how accurate signal matching is and does not provide a judgement on the model; in other words, it corresponds to a likelihood function, rather than an approximation of the evidence. Its normalization by the number of ``degrees of freedom'', referred to as \emph{reduced-$\chi^2$}, is a deceivingly simple and often unjustifiable metric for model selection \cite{Andrae2010DosChi-squared}. 

\subsection{Signal modeling and combination}
\label{sec:sig_modeling}
Every diagnostic has finite spatio-temporal resolution, thus limiting the accuracy and precision of measurements. In attempting to match \texttt{pySTRAHL} dynamics and experimental data, we integrate simulation results over time bins representing each instrument's photon collection time. Moreover, our inferences include a few useful \emph{nuisance} parameters, i.e. free variables that are not of interest \emph{per se}, but which are important in order to reproduce the physical dynamics of impurity injections. For example, we infer corrections to the estimated location of the sawtooth mixing radius and diagnostic time bases (which may not be exactly synchronized otherwise). We also allow the locations of spline knots for $D$, $v$ profiles to be freely inferred, forcing a small minimum distance between them to prevent exploration of extreme gradients of transport coefficients, which can lead to numerical instabilities. No evidence is observed of Ca recycling and therefore we avoid free parameters corresponding to this effect. 

Inspired by the astrophysical literature (e.g. Refs.\cite{Trotta2008BayesCosmology,Lahav2000BayesianMeasurements,Hobson2002CombiningEvidence,Marshall2006BayesianDatasets}), we adopt a Gaussian likelihood over measurement errors and reconsider the common practice\cite{Howard2012QuantitativePlasma, Odstrcil_2020} of weighting $\chi^2$ from different diagnostics with fixed factors as
\begin{equation}
    \chi^2 = \sum_k \alpha_k \chi_k^2
\end{equation}
where $k$ is an index identifying an experimental signal to be matched. The $\alpha_k$ weight is unknown, and commonly fixed by an experimentalist to match expectations on which signal matching should be prioritized. 
We consider this in a Bayesian light and look for an appropriate prior over $\alpha_k$. By a maximum-entropy argument (see Appendix~\ref{sec:appendixB} for details), one may find that an appropriate prior with unit expectation is the exponential $P(\alpha_k) = \exp(-\alpha_k)$.
In practice, it is often preferable to limit the prior to have a finite width, in order to avoid unreasonable values. While no analytic solution to this constrained entropy maximization is available\cite{Hobson2002CombiningEvidence}, the gamma distribution (of which the exponential is a special case) fulfills our objectives, while not strictly being a solution to the above problem. As with the exponential prior, we are able to integrate a Gaussian likelihood over such a gamma prior for $\alpha_k$ analytically (see Appendix~\ref{sec:appendixB}). This allows us to effectively substitute the simple $\chi^2$ metric with 
\begin{equation} \label{final3}
    \ln P(D|\theta) = \sum_{k=1} \left[ \ln \Gamma\left(\frac{n_k}{2}+\nu\right) - \left(\frac{n_k}{2}+ \nu\right) \ln \left(\frac{\chi_k^2}{2} +\nu \right) - \ln \Gamma(\nu) + \nu \ln \nu \right].
\end{equation}
where we defined $\nu \coloneqq a=1/b$ to fix the gamma prior mean $\langle \alpha_k \rangle=1$ and we dropped constant factors that do not depend on the inference parameters. This likelihood is analogous to the typical $\chi^2$ one, but sums over diagnostic weights that follow a gamma probability distribution with $\nu \vcentcolon= a=1/b$. 
By setting $\nu$ to different values, we allow $\alpha_k$ to be more or less free to weight diagnostics differently. Appendix~\ref{sec:appendixB} shows some possible choices of $\nu$ that we considered; the inference shown in section~\ref{sec:results} used $\nu=25$. Note that there is one value of $\alpha_k$ for each diagnostic $k$, but since we analytically marginalized the likelihood over the $\alpha_k$ prior there is no direct sampling of $\alpha_k$ to be done: inclusion of these parameters in an inference does not incur into any new free parameters (and thus adds no additional computational cost). 

By assigning a prior to the ``weight factors'', we allow our algorithm to infer appropriate values based on observed under-estimation of uncertainties, inaccuracies of atomic data, correlated signals, and other uncontrollable issues. Setting $\nu=25$ in Eq.~\ref{final3}, as we do in this work, conservatively constraints weights not be different from their first estimates by more than $\approx 50\%$.


\subsection{Nested Sampling} \label{sec:ns}
The \texttt{BITE} framework makes use of the Nested Sampling (NS) Monte Carlo method\cite{Skilling2004NestedSampling} for model selection and parameter estimation. The primary objective of NS is to evaluate the evidence integral in Eq.~\ref{eq:evidence}, which can be re-written as 
\begin{equation} \label{eq:NS_evidence_1}
    \mathcal{Z} = \int \mathcal{L} \ dX
\end{equation}
where $X$ is the prior survival function (referred to as ``prior volume'') above a given value of likelihood $\mathcal{L}$:
\begin{equation}
    X(\lambda) = \int_{\{\theta: \mathcal{L}>\lambda\}} \pi(\theta) d\theta. 
\end{equation}
$X(\lambda)$ may be thought of as the fraction of the prior hyper-volume in which $\mathcal{L}>\lambda$. By writing the evidence integral in the form of Eq.~\ref{eq:NS_evidence_1}, one reduces a complex multi-dimensional integral to a 1D integral, which may be computed via standard numerical methods once samples of $\mathcal{L}$ and $X$ have been collected. 

NS sets out to do so by first obtaining pseudo-random samples within a unit-hypercube having as many dimensions as the inference problem of interest. These samples are passed to a user-defined function that must transform them to the prior space of interest, usually via the inverse Cumulative Distribution Function (CDF) of the specific prior used for each dimension. 
Parameters are then passed to user-defined routines that run the forward model (in our case, \texttt{pySTRAHL} and synthetic diagnostics) and return a likelihood value. The NS algorithm thus collects the prior and likelihood probabilities for each unit-hypercube sample and orchestrates where future samples should be obtained in order to explore the (unknown) posterior distribution. NS makes use of multiple \emph{live points} that can explore the posterior in parallel. At any iteration, the live point with the lowest likelihood is eliminated from the set (becoming ``inactive''), and replaced with a new sample, which is accepted (becoming ``active'') if its likelihood is higher than the one of the eliminated point, or else is rejected. The algorithm therefore progressively explores nested shells of likelihood, reducing the prior volume that is sampled to find the best fitting parameters. Following this sampling procedure, the algorithm ``ranks'' the obtained values of likelihood based on their prior volume, $X$. It then becomes possible to evaluate the 1D integral in Eq.~\ref{eq:NS_evidence_1}, once a tolerance condition for the $\mathcal{Z}$ estimate is reached. As an important by-product of this process, NS allows one to use the set of live points 
from all iterations to reconstruct the explored posterior, possibly recovering any multimodal structure (i.e. separate peaks of the posterior distribution). In the ``Importance Nested Sampling'' (INS) variant, live points that were discarded at every iteration are also used to further improve the estimate of $\mathcal{Z}$ \cite{Feroz2013ImportanceAlgorithm}. 

The main difficulty in implementing the NS algorithm is the necessity to efficiently draw unbiased samples within iso-likelihood contours in the prior space. The \texttt{MultiNest} algorithm \cite{Feroz2008MultimodalAnalyses,Feroz2013ImportanceAlgorithm} does so by fitting (potentially overlapping) ellipsoids to the set of live points and sampling from within their union. Details of the algorithm and \texttt{MultiNest} parameters that we choose for transport inferences are given in Appendix~\ref{sec:Appendix_MultiNest}.

\subsection{Choice of priors}
\label{sec:priors}
As mentioned above, nested sampling is an inherently Bayesian method. The use of priors, on top of likelihoods, allows us to provide valuable information to the algorithm and avoid unphysical regions, while also demanding care to avoid inappropriate bias. \emph{We adopt the philosophy that all information available to us and not already encapsulated in the likelihood should be carefully put to use via priors}. Our choices are described in this section. 

Whenever possible, we avoid the use of uniform priors over a parameter value. These would imply that we have no idea of what is the most likely region of parameter space where true physical solutions lie - something that is often not true. For example, we apply a Gaussian prior, rather than a uniform one, over the sawtooth inversion radius, estimated with $\sim 1$ $cm$ spatial accuracy via ECE or SXR diagnostics. Adopting a uniform prior over $D$ would also be a mistake, given that we are interested in exploring diffusion that may go from neoclassical expectations near the magnetic axis (of the order of $10^{-2}$ $m^2/s$) to much larger values, possibly approaching $100$ $m^2/s$ at mid-radius, where turbulent transport is dominant. \emph{Uninformative} sampling in this case requires one to apply uniform sampling to $\log(D)$; in the Bayesian literature, this is referred to as a \emph{Jeffreys prior}\cite{Jeffreys1946AnProblems.,Jaynes1968PriorProbabilities}.

Assuming that neoclassical and turbulent processes act in such a way as to eliminate fine structure from radial profiles, we expect $D$ and $v$ profiles to vary smoothly as a function of radius. In order to encapsulate this expectation into our priors we make use of \emph{Gaussian copulae}, through which we set correlations between the values of $D$ and $v$ (separately) at adjacent spline knots. A Gaussian copula ``couples'' different sampled parameters, $\vec{u}=[u_1,u_2,\dots, u_d]$, from a $d$-dimensional unit hypercube. It is defined via the expression
\begin{equation} \label{eq:copula}
    C_R^{Gauss}(\vec{u}) \coloneqq \Phi_R \left(\Phi^{-1}(u_1), \dots, \Phi^{-1}(u_d) \right) 
\end{equation}
where $R$ is a correlation matrix, $\Phi_R$ is the joint cumulative distribution function of a multivariate normal distribution with zero mean vector and covariance matrix $R$, and $\Phi^{-1}$ is the inverse cumulative distribution function of a standard normal. The result of the transformations in Eq.~\ref{eq:copula}, $C_R^{Gauss}(\vec{u})$, is a set of parameters that are individually (i.e. marginally) uniformly distributed, but not \emph{jointly} uniformly distributed in the unit hypercube. Rather, they present correlations indicated by the correlation matrix $R$. We choose $R$ to be tridiagonal, with diagonal values of $1.0$ and off-diagonal entries of $0.5$. This makes our expectation of ``smoothness'' in $D$ and $v$ profiles more explicit without necessarily forcing personal bias on the posterior \cite{Nelsen2007Introduction}. 

We choose to apply prior constraints over $D$, $v/D$ rather than $D$, $v$. We find this to be more conservative since $v/D$ is related to density peaking and is unlikely to take very large values; on the other hand, $v$ may vary widely so long as a physical $v/D$ is maintained (see section~\ref{sec:tglf}). 
In devices with direct impurity density measurements (e.g. via CER), values of $v/D$ may be strongly constrained; however, on C-Mod only line-integrated brightness is available and this appears to constrain $D$ more strongly than $v/D$. We therefore impose a relatively restrictive prior on $v/D$ [$m^{-1}$] at midradius, using a Gaussian distribution with mean of 0 and standard deviation of $2$ $m^{-1}$ ($\mathcal{N}(0,2)$). This allows enough freedom to explore positive and negative peaking within a range that spans measurements on other devices \cite{Angioni2011GyrokineticUpgrade,Odstrcil_2020,Manas2017GyrokineticPlasmas, Villegas2014ExperimentalSupra} and encompassing observed values of $1/L_{n_e}\sim 1$ at midradius in the I-mode discharge described in this paper. Near-axis, we allow significantly more variation with $v/D\sim\mathcal{N}(0,10)$. In the pedestal, where we have only weak experimental constraints but one may expect large inward pinch convection, we set weak priors on the Gaussian pinch amplitude ($\sim\mathcal{N}(-100,50)$ ) and its width ($\sim\mathcal{N}(0.03,0.03)$, in $r/a$ units)\footnote{Truncated normal priors are used when appropriate to prevent sampling of unphysical parameters, e.g. negative radial ``widths''.}. 
We also impose that a ``non-negligible'' fraction of LBO-injected particles should enter the confined plasma in \texttt{pySTRAHL} simulations. Previous work (e.g. Ref. \cite{Rice1997XPlasmas}) reported LBO penetration fractions equal or greater than 5-10\%, suggesting that one may conservatively set a minimum penetration fraction of 1-2\% in \texttt{BITE}. This simple condition makes a large $D$ ($\gtrsim 5$ $m^2/s$) and a small inward edge $v$ ($|v| \lesssim 50$ $m/s$) at the LCFS likely unrealistic.
Finally, in the absence of more detailed knowledge, time base synchronization of experimental signals, being determined by random triggering events in each detector, is set via uniform priors with appropriate bounds for each detector.

%
%
%
%
%
\section{Experimental Results} \label{sec:results}

The inference of $D$, $v$ profiles and their uncertainty is strongly dependent on the flexibility of the $D$ and $v$ ``models'' that one applies. In this section, we illustrate in turn the results of model selection and parameter estimation for the I-mode case of interest. 


\subsection{Model Selection}
\begin{figure}[ht]
	\centering
	\includegraphics[width=0.5\textwidth]{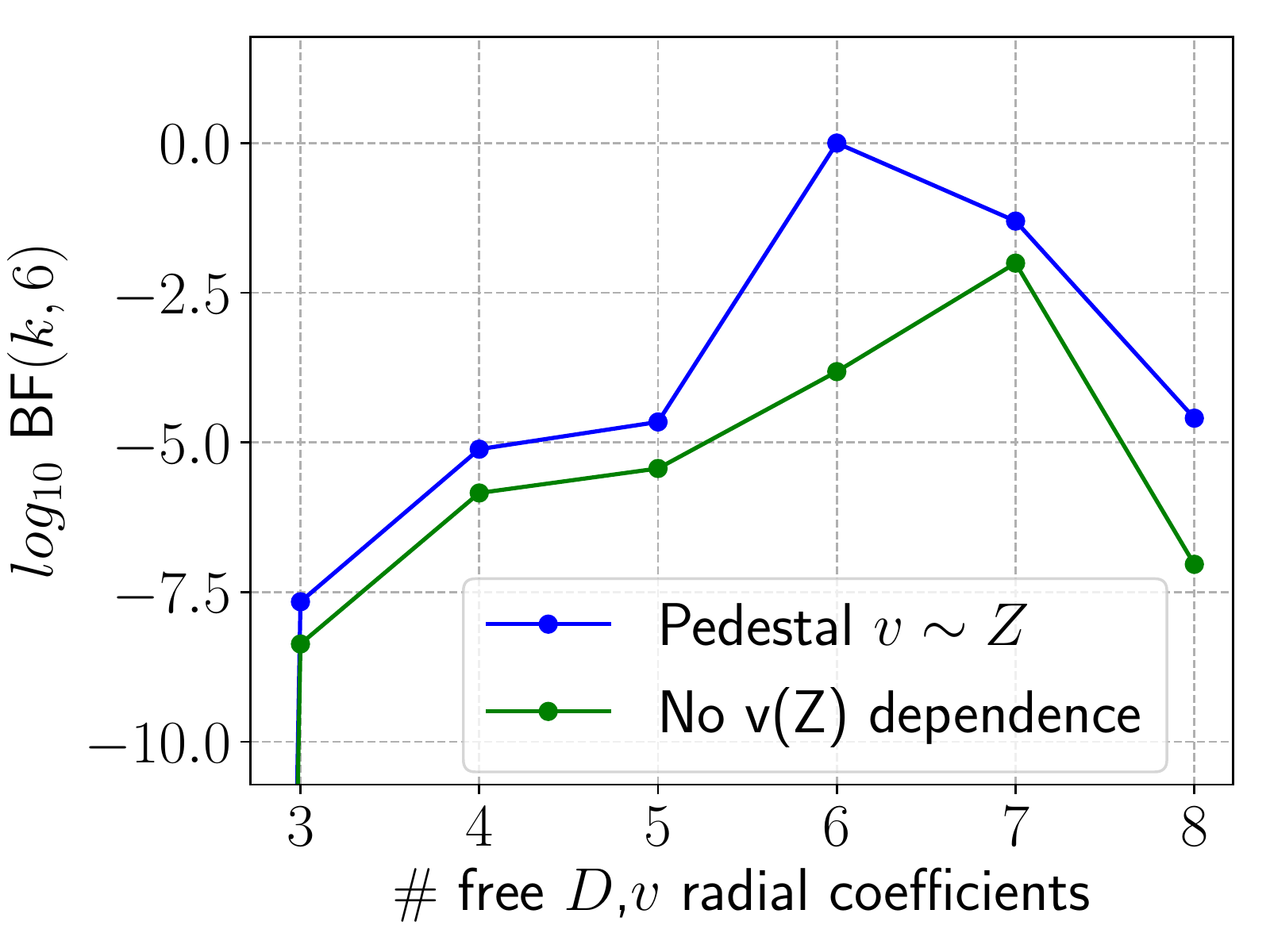}   
	\caption{Scaling of the log of the Bayes Factor (BF) with number of free $D$, $v$ radial coefficients in the analyzed I-mode discharge. The model with 6 free $D$ and $v$ radial coefficients and with $v$ scaling with $Z$ in the pedestal is consistently, although weakly, favored by experimental data at all dimensionalities.}
	\label{fig:lnev_scaling}
\end{figure}

As described in section~\ref{sec:bayesian}, the Bayesian evidence is the central metric in Bayesian inference to assess what level of model complexity is best supported by data. The Bayesian evidence may be interpreted as a \emph{relative} metric, since the ratio of values of evidence have a rigorous statistical interpretation for model selection (see Ref.\cite{Trotta2008BayesCosmology} for a detailed discussion). In Fig.~\ref{fig:lnev_scaling}, we show the base-10 logarithm of Bayes Factors (BF), defined as ratios of Bayesian evidence values, for a range of model complexities (number of free parameters). Each BF is relative to the case with highest log-evidence. Cases on the left of the BF peak are under-fitting, i.e. their models are insufficient to represent experimental data adequately; cases on the right are over-fitting, i.e. they require more parameters than experimental uncertainties suggest is reasonable to use. In all cases, we set $D$ and $v$ radial knots to be the same for simplicity. 

We compare the BF obtained when we set $v$ to scale linearly with $Z$ in the pedestal (taken as $r/a>0.9$) in \texttt{pySTRAHL} (blue) and when we set $v$ to be independent of $Z$ (green). From this plot, it is clear that the $v\sim Z$ model is best supported by data, although the strength of statistical evidence favoring the model-selected case with respect to the next case with the highest BF is relatively weak. However, the difference in predicted posterior distributions between these cases is also small, thus making the distinction between them not particularly significant. On the other hand, the difference between posteriors of cases differing by more than 3 units of $log_{10}(BF)$ can be dramatic. None of the cases with high BF were found to be multimodal, i.e. having separate statistical modes of comparable evidence, although the posterior distributions often exhibit non-Gaussian features.   

Obviously, uncertainties on experimental data strongly affect the dimensionality achieving the highest log-evidence (or, equivalently, BF). The importance of carefully quantifying such uncertainties for challenging model selection problems such as ours cannot be overstated; our detailed approach was described in previous sections. 



\subsection{Parameter Estimation}
\begin{figure}[ht]
	\centering
	\includegraphics[width=0.7\textwidth]{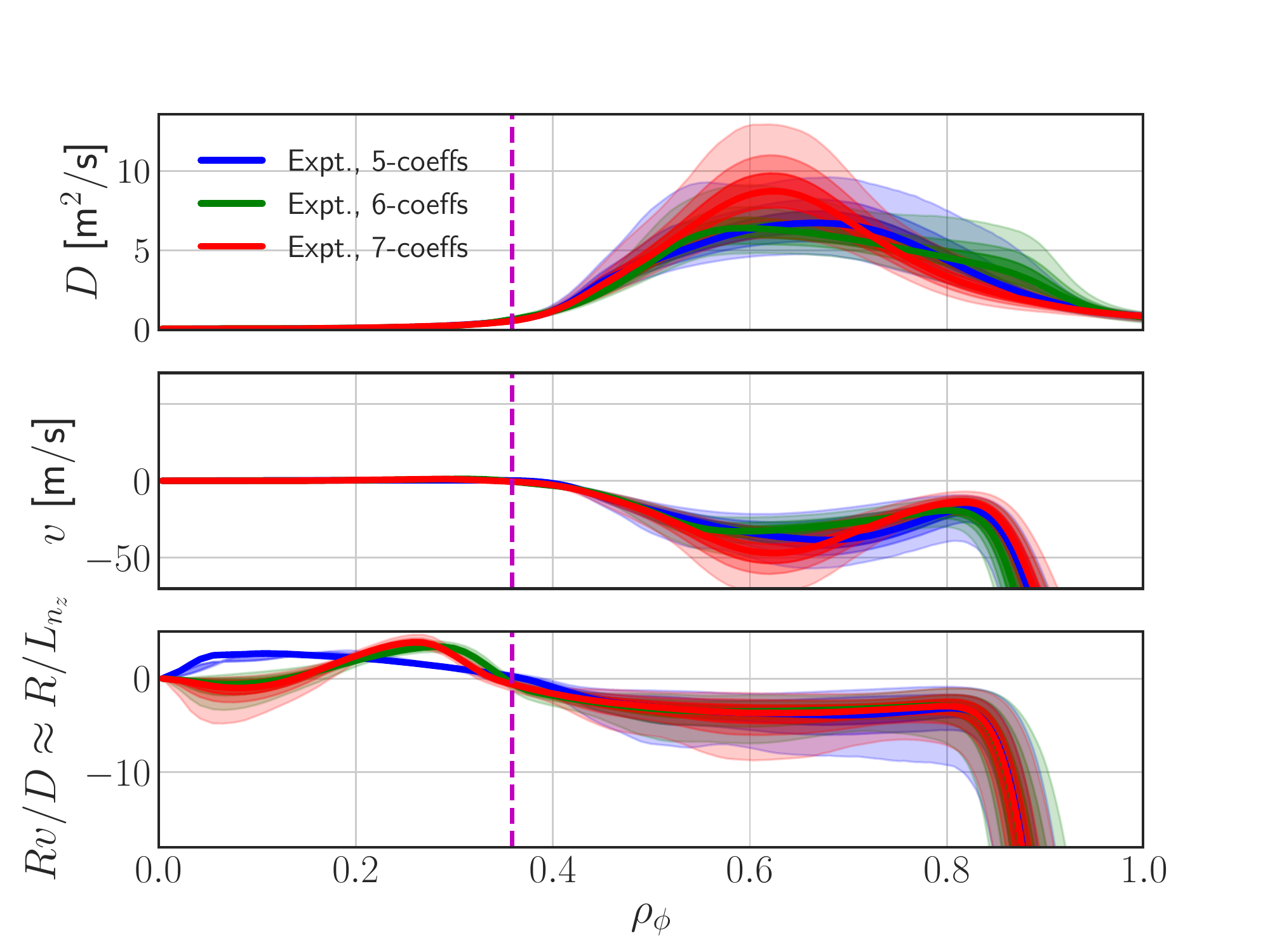}
	\caption{Inferred profiles of $D$ and $v$, together with the $R v/D$ ratio, for inferences with 5, 6 and 7 radial $D$ and $v/D$ coefficients (with $v$ calculated from their product). The 6-coefficient inference has higher log-evidence and is therefore \emph{model selected}; 5- and 7-coefficient cases are only shown for reference.} 
	\label{fig:imode_DVprofs_final1}
\end{figure}

Fig.~\ref{fig:imode_DVprofs_final1} shows the inferred $D$, $v$ and $v/D$ profiles from inferences obtained with the three levels of complexity that resulted in the highest BF (cf. Fig.~\ref{fig:lnev_scaling}), with 5 (blue), 6 (green) and 7 (red) radial coefficients for $D$ and $v$. All of these cases used $v(r/a>0.9)\sim Z$, since this choice is found to give higher BF. In Fig.~\ref{fig:imode_DVprofs_final1} we display profiles up to the LCFS, showing that a large inward pinch (plotted for fully-stripped Ca) is inferred in the pedestal. The maximum magnitude of such pinch ($\approx -150$ $m/s$) is not shown in order to visualize both core and edge profiles on the same figure; its value is not very well constrained by our data and depends on details of the applied model. The apparent existence of a pedestal impurity pinch does not preclude a short impurity confinement time: in this discharge, $\tau_{imp} \approx 30\pm2$ ms, while the energy confinement time $\tau_e\approx 23\pm1$ ms. 

The vertical magenta dashed line show the approximate location of the sawteeth inversion radius. Sample signal fits resulting from these transport coefficients are shown in Fig.~\ref{fig:signals_decays} and Appendix B. We remark that by setting the w and z line signals to be relatively calibrated, we make their fitting much more challenging: virtually no discrepancy between simulated and experimental signals would be seen if we allowed lines to be independently normalized\footnote{This effect is \textbf{not} due to inaccurate atomic rates, but to true physical complexity of experimental data.}. The uncertainties shown in Fig.~\ref{fig:imode_DVprofs_final1} are not assumed to be Gaussian and represent the 1-99, 10-90 and 25-75 quantiles of the posterior function. The Bayesian weight factors described in section~\ref{sec:sig_modeling} (see also Appendix ~\ref{sec:appendixB}) are found to effectively weight XICS signals (for both the w and z lines) more heavily than VUV lines. All signals are found to have effective weights of less than 1, indicating that diagnostic uncertainties are likely too small to have full consistency with the \texttt{pySTRAHL} model. The use of Bayesian weight factors results in an increase of error bars for the profiles in Fig.~\ref{fig:imode_DVprofs_final1} by approximately $30\%$ at $\rho_{\phi}=0.6$, compared to the results obtained with fixed diagnostic weights of 1. Setting $v\sim Z$ in the pedestal is found to lower the absolute magnitude of edge values of $D$ and $v$ required to match data, without significantly affecting inferred profiles in the core. 

\begin{figure}[ht]
	\centering
	\includegraphics[width=0.7\textwidth]{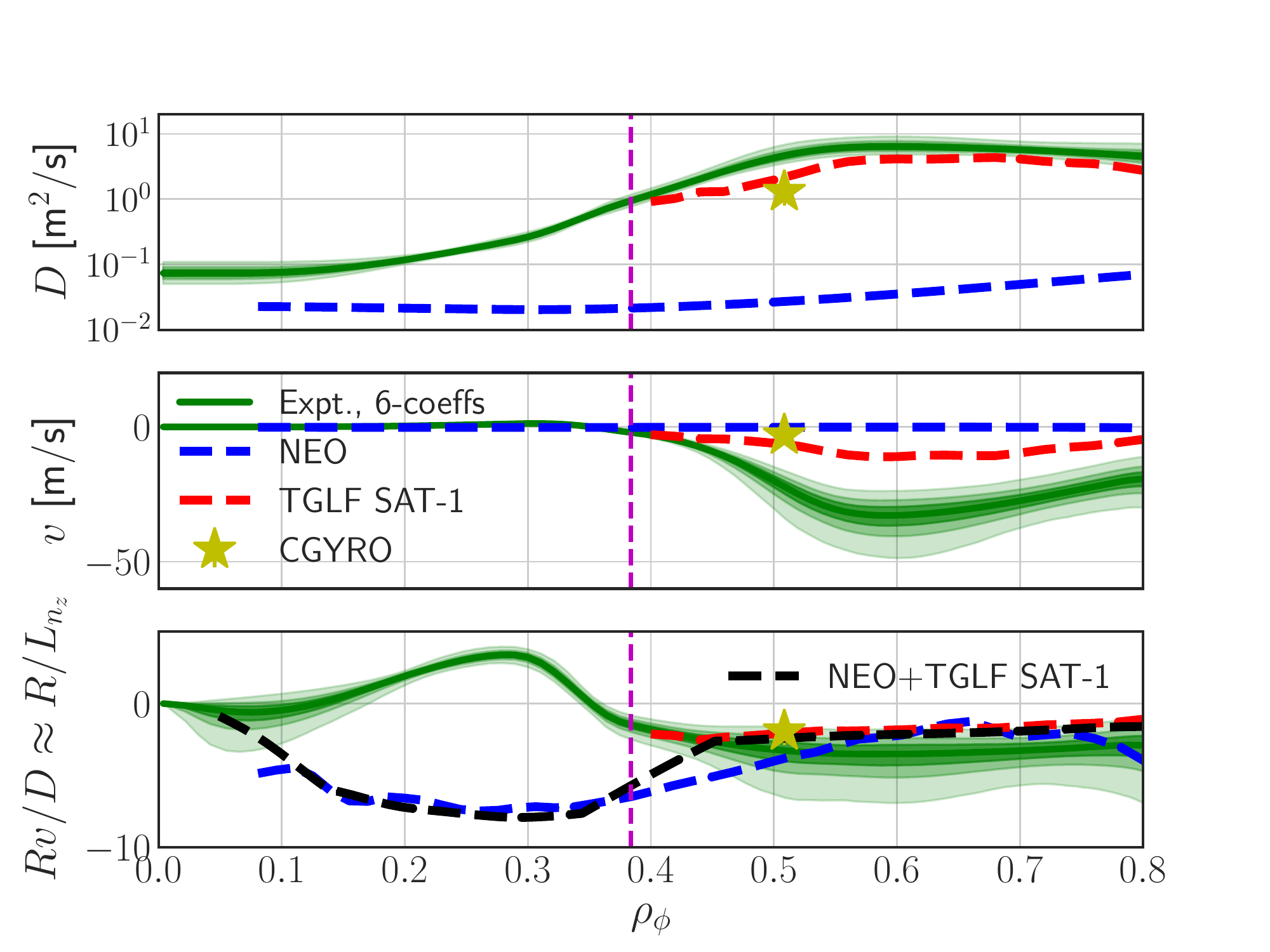} 
	\caption{Core-focused view of inferred profiles of $D$ (on a logarithmic scale) and $v$, together with the $R v/D$ ratio, for the model-selected inference (also shown in Fig.~\ref{fig:imode_DVprofs_final1}. Neoclassical, gyro-fluid and gyrokinetic modeling are overlaid. The black line in the lowest plot shows an interpolated profile of combined NEO and TGLF predictions for the total $R$ $v/D$.} 
	\label{fig:DV_inference_w_models_042020}
\end{figure}

On top of the experimentally-inferred radial profiles of impurity transport coefficients for the model with the highest BF, Fig.~\ref{fig:DV_inference_w_models_042020} shows the results of selected transport models. In blue, we show predictions from the NEO\cite{Belli2008KineticDynamics, Belli2012FullSimulations} neoclassical transport code and in red from the Trapped Gyro-Landau Fluid (TGLF) reduced quasilinear turbulence model\cite{Staebler2007APhysics,Staebler2017ATurbulence}. In the bottom panel, we also show the prediction for the total $R v/D$ value (black), taken as an additive measure of neoclassical and turbulent transport (i.e. $R \ v/D_{tot} = R(v_{neo}+v_{turb})/(D_{neo}+D_{turb})$). Scales have been adapted in each panel to complement the visualization of Fig.~\ref{fig:imode_DVprofs_final1} and allow better comparison to transport models.

NEO is a multi-species drift-kinetic code that solves the first-order (in the drift-ordering parameter $\rho^* = \rho_i/a \ll 1$) drift-kinetic-Poisson equations \cite{Belli2008KineticDynamics, Belli2012FullSimulations}. Since neoclassical predictions are formally limited by near-axis potato orbits and by ion orbit losses near the LCFS \cite{Belli2012FullSimulations,Helander2010CollisionalPlasmas}, we limit our NEO simulations to the range $0.1 \lesssim \rho_{\phi} \lesssim 0.95$. 
Fig.~\ref{fig:DV_inference_w_models_042020} shows that inferred diffusion values near the magnetic axis are larger, but of the same order of magnitude, as the NEO predictions. The inferred values are, of course, to some extent dependent on the applied sawtooth model in \texttt{pySTRAHL} simulations, but we have found that $D$ and $v$ profiles of neoclassical character are always inferred inside the sawtooth inversion radius. We note that this is only an approximate comparison, given that NEO was run with time-averaged kinetic profiles, as opposed to the time-dependent \texttt{pySTRAHL} predictions including multiple impurity charge states and sawtooth modeling. The NEO profiles displayed here were obtained for Ca$^{18+}$ (the dominant charge state in this discharge); future work will present more detailed predictions accounting for the presence of multiple charge states.


At mid-radius, where the effect of sawteeth is reduced, TGLF SAT-1 compares favorably with absolute magnitudes of turbulent $D$, but differs by more than a factor of 2 from $v$ (and $v/D$) experimental estimates. Fig.~\ref{fig:imode_DVprofs_final1} also shows a single-point prediction for $D$, $v$ and their ratio $v/D$ from a nonlinear, ion-scale, $Q_i$-matched CGYRO\cite{Candy2016APlasmas} simulation at $r/a=0.6$. 
Both TGLF and CGYRO modeling are described in greater detail in the next section. 

\begin{figure}[ht]
\centering
\begin{subfigure}{0.5\textwidth}
  \centering
  \includegraphics[width=\linewidth]{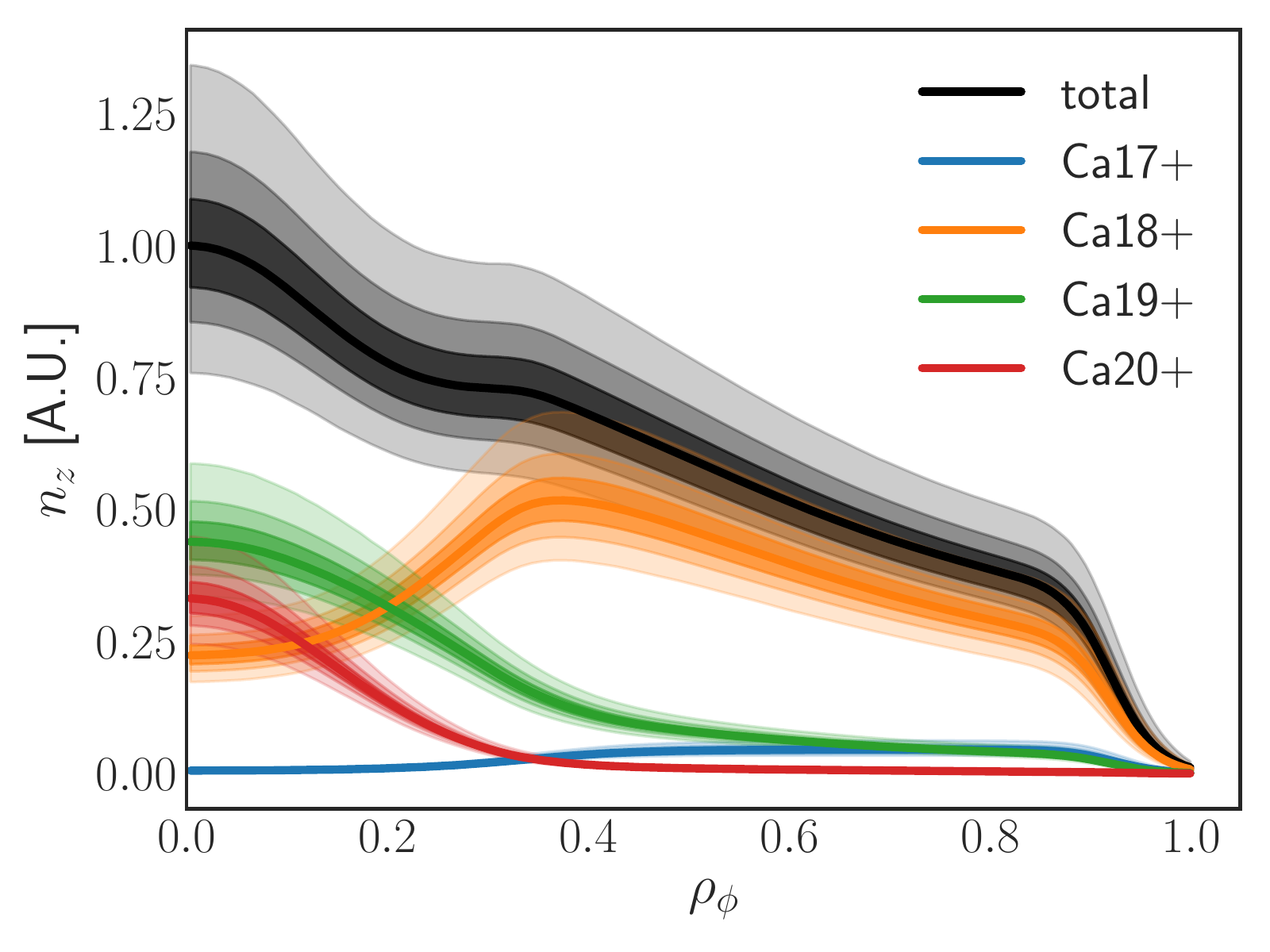}
  \caption{}
  \label{fig:nz_profs_imode}
\end{subfigure}%
\begin{subfigure}{.5\textwidth}
  \centering
  \includegraphics[width=\linewidth]{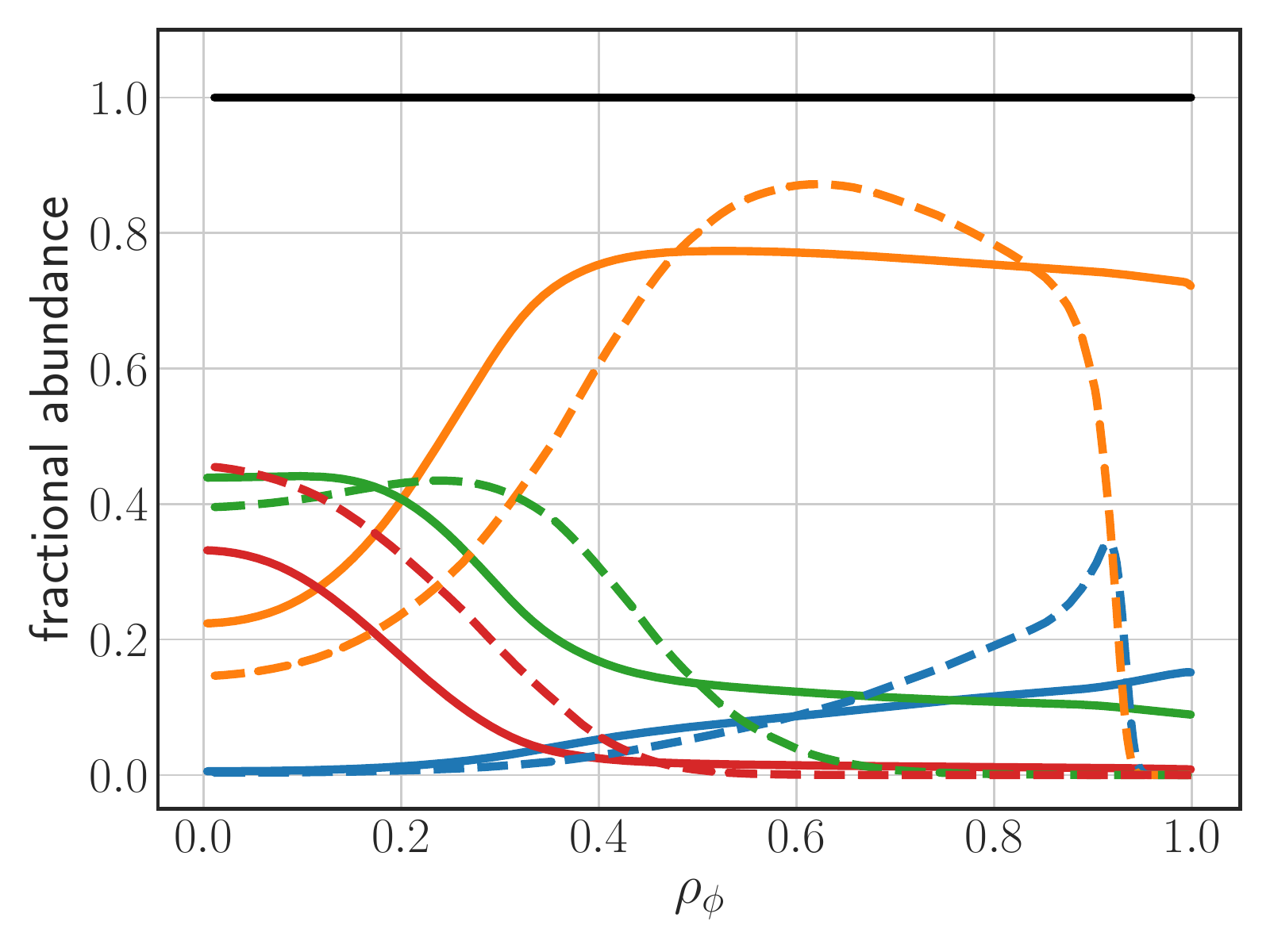}
  \caption{}
  \label{fig:frac_profs_imode}
\end{subfigure}
\caption{(a) Experimentally-inferred self-decaying profile shapes for the highest ionization stages of Ca after an LBO injection into the I-mode discharge. In (b), we compare the median fractional abundances of (a), shown with continuous lines, with those predicted by ionization equilibrium only (no transport), shown with dashed lines.}
\label{fig:test}
\end{figure}

In Fig.~\ref{fig:nz_profs_imode} we show the self-similar impurity density profiles for the highest ionization stages corresponding to the inferred transport coefficients in Fig.~\ref{fig:DV_inference_w_models_042020}, normalized such that the median of the total density is 1 on axis. By ``self-similar'' here we refer to the property of maintaining profile shape while decreasing in magnitude over time after the LBO injection.
Density uncertainties were computed from the inferred $D$, $v$ uncertainties, i.e. by propagating $D$, $v$ uncertainties using chains and weights from \texttt{MultiNest}. This gives a clear visualization of how uncertainties in Fig.~\ref{fig:DV_inference_w_models_042020} translate to $n_z$ profiles, retaining all degrees of correlation between parameters. 

In Fig.~\ref{fig:frac_profs_imode} we compare the median fractional abundances for the highest charge states, shown with continuous lines, with those given by only balancing effective ionization and recombination rates, i.e. without any transport. This figure shows that the presence of transport significantly modifies the localization of each charge state. In particular, we note that it allows these highly-ionized charge states to live in the pedestal region, where atomic physics by itself would predict that only lower charge states should exist. This excess of highly-ionized states in the pedestal is not present during the rise phase of the LBO injection, since impurity ions must first penetrate into the core to undergo nearly-complete ionization and then be transported outwards; only at this stage, the upper level of x-ray transitions can be significantly populated via recombination, as previously described by Rice \emph{et al.}~\cite{Rice1995X-rayTokamak}. Such recombination effect contributes to the He-like Ca z line emissivity measured via XICS, but not to the corresponding w line, as described in section~\ref{sec:spectroscopy_methods}, thus providing a non-trivial constraint to our inference.

\section{Turbulence Modeling} \label{sec:modeling}
Having described our attempts at quantifying transport coefficients for the I-mode case of interest, we now proceed to describe in greater depth the turbulence modeling to which these results can be compared. The turbulence modeling in this section is not in any way informed by the experimental results presented above, thus offering the means for an independent validation effort.

In this section, we describe both nonlinear gyrokinetic simulations and quasilinear modeling. The latter is commonly used for integrated modeling, where fast iterations are required to interpret experimental observations or predict plasma behavior. In particular, we make use of TGLF\cite{Staebler2007APhysics,Staebler2017ATurbulence}, which solves the Trapped Gyro-Landau Fluid equations to approximate the linear eigenmodes of gyrokinetic drift-wave instabilities: trapped ion and electron modes (TIM, TEM), ion and electron temperature gradient modes (ITG, ETG), and kinetic ballooning modes (KBM). Using these, TGLF computes quasilinear estimates to turbulent fields\cite{Staebler2005Gyro-LandauParticles, Staebler2007APhysics}, making use of a parametric saturation rule whose coefficients are set to match a database of fully-nonlinear gyrokinetic simulations. 



By adding trace Ca impurities to TGLF or CGYRO runs, one may obtain estimates for turbulent diffusion and convection components. This is a valid procedure because \emph{in the trace limit} a species has such a low concentration as to become negligible in the Poisson equation and in Amp\`ere's Law. Consequently, the turbulent electromagnetic potential becomes independent of the distribution function of the trace particles and the particle flux becomes \emph{linear} in the thermodynamic gradients\cite{Angioni2009GyrokineticModelling}:
\begin{equation} \label{eq:theory_fluxes}
\frac{R \Gamma_{z}}{n_z} =  D_z \frac{R}{L_{n,z}} + D_{T,z} \frac{R}{L_{T,z}} + R v_{p,z}
\end{equation}
Here we separated the effect of density and temperature gradients on the impurity flux, and defined a particle diffusion coefficient $D_z$ (identical in definition to what we obtained in our experimental analysis), a thermodiffusion term $D_{T,z}$ and a pure convection term $v_{p,z}$ ($z$ subscripts will be dropped henceforth). For clarity, we also define $v_{T,z} \coloneqq D_{T,z}/L_{T,z}$, which has units of m/s and is more properly identified as a thermal convection term. Some authors also define a term that is linearly proportional to toroidal rotation gradients (``roto-diffusion''), e.g. Refs.\cite{Angioni2017APlasmas,Casson2010GyrokineticPlasma}. We shall not consider this since having species with different toroidal rotation is not possible in CGYRO and this prevents us from rigorously studying this term. Roto-diffusion is thus ``grouped'' in this analysis with other convection terms. 

In order to get $D$, $v_T$ and $v_p$ estimates from TGLF or CGYRO, we introduce 3 trace species. These may either be all included in the same run, as we do for CGYRO, or in separate runs; the latter is a better option for codes like TGLF whose computational cost scales with the square of the number of species. Relative to the main-ion density, we slightly vary\footnote{The actual percentage variation is inconsequential for trace-level concentrations.} the input density gradient of the second and third species and, similarly, the input temperature gradient of the third species only. We then collect a set of linear equations for each species' flux into matrix form
\begin{equation} \label{eq:matrix}
 \frac{R}{n_z}  \begin{pmatrix} \Gamma_{z,1} \\  \Gamma_{z,2} \\  \Gamma_{z,3} \end{pmatrix} = \begin{pmatrix}
R/L_{n,z,1} & R/L_{T,z,1} & 1\\
R/L_{n,z,2} & R/L_{T,z,2} & 1 \\
R/L_{n,z,3} & R/L_{T,z,3} & 1
\end{pmatrix} 
\begin{pmatrix} D_{z} \\  v_{T,z} \\  v_{p,z} \end{pmatrix} 
\end{equation}
By inverting the $3\times 3$ matrix, one can then solve for the local transport coefficients. An analogous method was also used to obtain the NEO and TGLF $D$ and $v$ radial profiles in Fig.~\ref{fig:DV_inference_w_models_042020}. In this case, we omitted the temperature perturbation in Eq.~\ref{eq:matrix} and thus only obtained a total convection term (labeled  $v$), since thermal convection cannot be separated experimentally by our inference framework. It is nonetheless interesting to distinguish $v_T$ from $v_p$ in theoretical modeling to compare to predictions for thermodiffusion in different turbulent regimes \cite{Angioni2012Off-diagonalAspects}.


\subsection{Quasilinear TGLF modeling} \label{sec:tglf}
In this section, we explore the sensitivity of the radial profiles of impurity transport coefficients predicted by TGLF, shown in Fig.~\ref{fig:DV_inference_w_models_042020}. We make use of the SAT-1 ``multiscale'' saturation rule \cite{Staebler2016TheTurbulence,Howard2016Multi-scaleTransport} since this is found to better reproduce kinetic profiles while matching experimental heat fluxes from TRANSP \cite{Breslau2018TRANSP} within the TGYRO framework \cite{Candy2009TokamakSimulation}.

Analysis of the TGLF linear spectrum shows that ion-scale modes with $k_y \rho_s$ up to $\approx 0.9$ have real frequencies indicating propagation in the ion diamagnetic drift direction. Here, $k_y=nq/r$, $n$,$q$ and $r$ are the toroidal mode number, safety factor and outboard midplane minor radius, respectively, and $\rho_s$ is the ion sound speed gyroradius. Together with a strong sensitivity to scans in $a/L_{T_i}$, and weaker sensitivity to other typical drift wave drives, linear spectra suggest that ITG modes are strongly dominant in this plasma, in agreement with past I-mode simulations~\cite{White2015NonlinearExperiment,Fu2013TurbulentC-Mod}. The observation of positive thermal convection, $v_T>0$, also confirms the identification of ITG being dominant\cite{Angioni2012Off-diagonalAspects}. While TGLF and CGYRO spectra are different, particularly at scales where ITG and TEM modes co-exist, they agree in the identification of the most intense turbulent mode being at $k_y \rho_s \approx 0.4$. 

\begin{figure}[ht]
	\centering
	\includegraphics[width=1.0\textwidth]{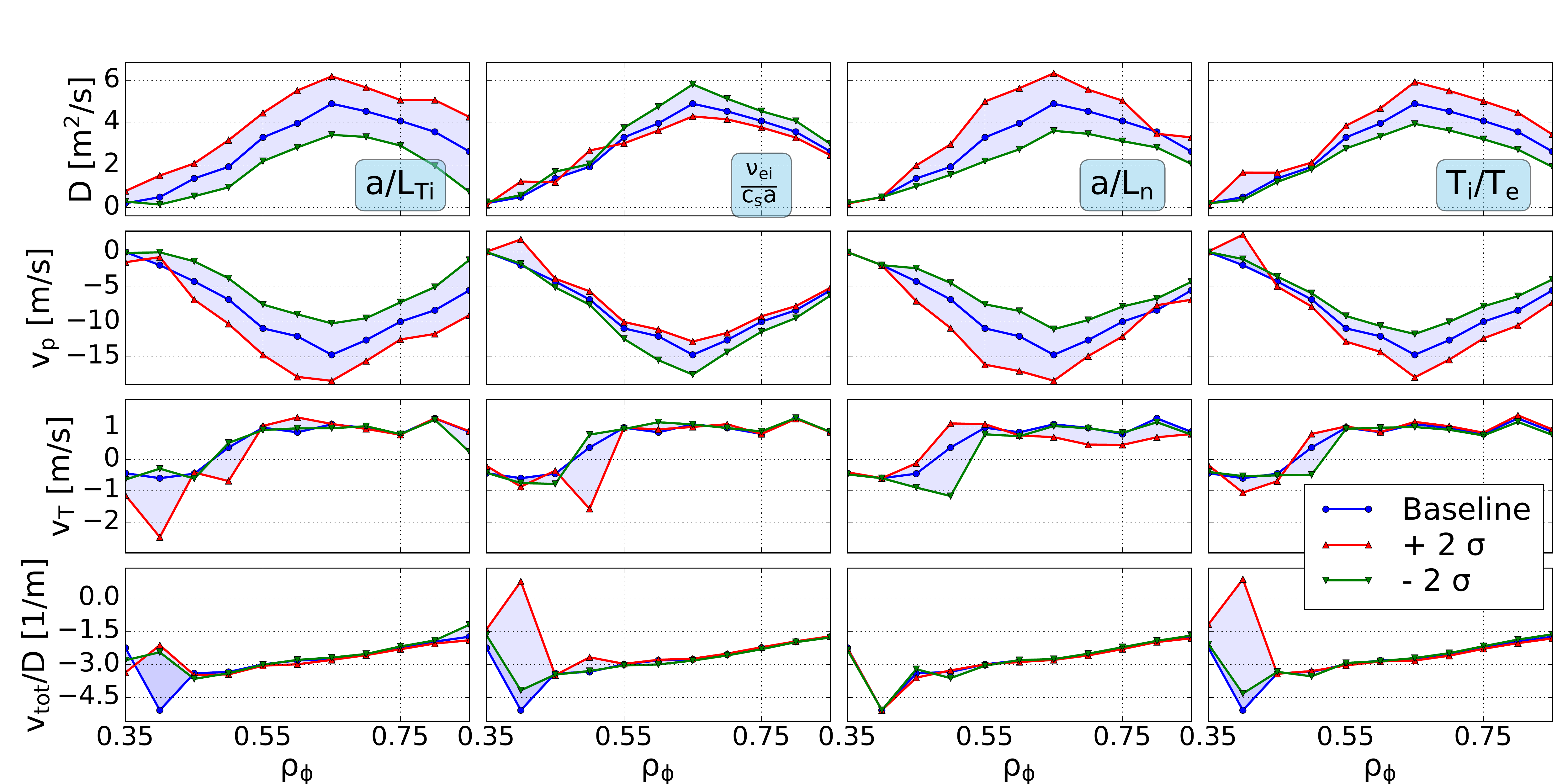} 
	\caption{Diffusion and convection coefficients from TGLF $2 \sigma$ scans around experimental I-mode parameters. Uncertainties are estimated locally at each radial location for each parameter.}
	\label{fig:Imode_DVscans_101318}
\end{figure}

In Fig.~\ref{fig:Imode_DVscans_101318} we show independent TGLF scans of $a/L_{T_i}$, $a\nu_{ei}/c_s$, $a/L_n$ and $T_i/T_e$ within $2$ standard deviations, as estimated from experimental data (cf. Fig.~\ref{fig:MITIM1_paper_kinetic_profiles}). Here, $c_s$ is the local sound speed and $\nu_{ei}$ is the electron-ion collision frequency, for which we considered an uncertainty of $20\%$ following Ref. \cite{Sung2016QuantitativeDischarges}. We use $2 \sigma$ rather than the more conventional $1 \sigma$ in order to encompass $\approx 95\%$ of possible outcomes, taking all uncertainties to be Gaussian. 
These scans suggest that $D$ and the total $v$ have an overall uncertainty of at least $30\%$ of their magnitude. We note, however, that a complete assessment of uncertainties and comparison to the \texttt{BITE} results in Fig.~\ref{fig:DV_inference_w_models_042020} would ideally require varying input parameters simultaneously; correlations across TGLF inputs could be important, but are not available from the 1D scans of Fig.~\ref{fig:Imode_DVscans_101318}. Interestingly, $v/D$ is almost fixed within the radial range that we explore; in other words, changes of $D$ and $v$ compensate such that the expected impurity peaking indicated by $v/D$ remains constant. The effect of parallel velocity shear on these results has been found to be negligible. Transport coefficients were found to generally have weaker dependence on $a/L_{T_e}$ than on $a/L_{T_i}$ (not shown). However, greater sensitivity to $a/L_{T_e}$ at outer radii suggests that electron-scale modes grow stronger (and ion-scale modes weaker) as one approaches the top of the pedestal; this is found to correlate with smaller diffusion and convection magnitudes.  


\subsection{Gyrokinetic modeling}
\label{sec:cgyro}
We next describe results from nonlinear ion-scale gyrokinetic simulations with CGYRO \cite{Candy2016APlasmas}. We focused our runs on $r/a=0.6$, where experimental profiles are well determined and the effect of sawteeth is expected to be small. We used a domain size with $L_x/\rho_s\approx 100$ and $L_y/\rho_s\approx 100$, radial grid spacing of $\Delta x/\rho_s= 0.061$ and $\Delta y/\rho_s= 0.065$, with 344 radial modes and 22 toroidal modes, giving $\max(k_x \rho_s) \approx 10.5$ and  $\max(k_y \rho_s) \approx 1.4$. 
In velocity space, we use a grid with 16 pitch angles, 8 energy and 24 poloidal points. We adopt experimental profile inputs, Miller geometry, electromagnetic ($\tilde{\phi}$ and $\tilde{A_{||}}$) effects, as well as gyrokinetic electrons. We apply the sonic rotation scheme recently added to CGYRO\cite{Belli2018ImpactTransport}, whose effects with respect to the linear ordering in Mach number are known to account for significant differences of transport for heavy impurities\cite{Casson2015TheoreticalUpgrade}. Averages and uncertainties on simulation outputs are estimated using a moving average over windows of length equal to 3 times the estimated correlation time. 


Sensitivity scans for each of the parameters mentioned above, as well as for numerical dissipation values, were run by increasing each parameter by $50\%$ in turn and checking for significant variations in output. The I-mode discharge of interest has a value of dimensionless electron collision frequency (defined as in Ref.\cite{Candy2016APlasmas}) of $\bar{\nu}_{ee}\approx 0.1$ at the examined location - a relatively \emph{low}-collisionality case for C-Mod. Linear scans of collisionality did not display strong dependencies in any outputs, but did highlight that this parameter affects particle transport more strongly than heat transport. This may not be surprising, in view of the well-known role of collisionality on electron transport and density peaking\cite{Angioni2003DensityPlasmas,Angioni2009ParticleExperiment,Greenwald2007DensityScalings}. Application of the Sugama operator \cite{Sugama2009LinearizedEquations, Belli2017ImplicationsSimulation,Candy2016APlasmas}, as opposed to pitch-angle scattering via the Lorentz operator, has also been found to strongly affect particle transport predictions. 

Nonlinear scans of $a/L_{T_i}$ were used to find the value at which CGYRO could match the experimental turbulent ion heat flux, $Q_i$, as determined by power balance via TRANSP\cite{Hawryluk1981ANTRANSPORT}. This was found to be at $a/L_{T_i}\approx1.65$, approximately $25\%$  below the experimental $a/L_{T_i}$ estimate, for which CGYRO predicts a value of $Q_i$ that is 6 times higher. As in previous GYRO simulations of I-mode discharges\cite{White2015NonlinearExperiment}, we have found this case to be particularly stiff and close to marginality. This makes an exact $Q_i$ matching procedure more difficult, since $Q_i$ varies significantly over time. Consistently with past work \cite{Mikkelsen2018VerificationC-mod, Creely2017ValidationC-Mod}, we find a clear underestimation of $Q_e$ at all scanned values of $a/L_{T_i}$. Changes in $a/L_{T_e}$ (by $10\%$) or in $E\times B$ shear (by $50\%$) saw no significant variation of $Q_e/Q_i$ (less than $10\%$), which remains less than $1/2$ of the value predicted by TRANSP. This is consistent with expectations from linear CGYRO scans, which show little sensitivity to typical TEM or ETG drives, suggesting that multiscale simulations \cite{Howard2016Multi-scaleTransport} may still be unable to resolve the observed discrepancies in $Q_e$. Nonetheless, given the predominant role of ion-scale fluctuations in determining turbulent impurity transport, these results do not preclude a useful comparison of impurity transport predictions with experiment. 

\begin{figure}
\centering
\begin{subfigure}{0.5\textwidth}
  \centering
  \includegraphics[width=.99\linewidth]{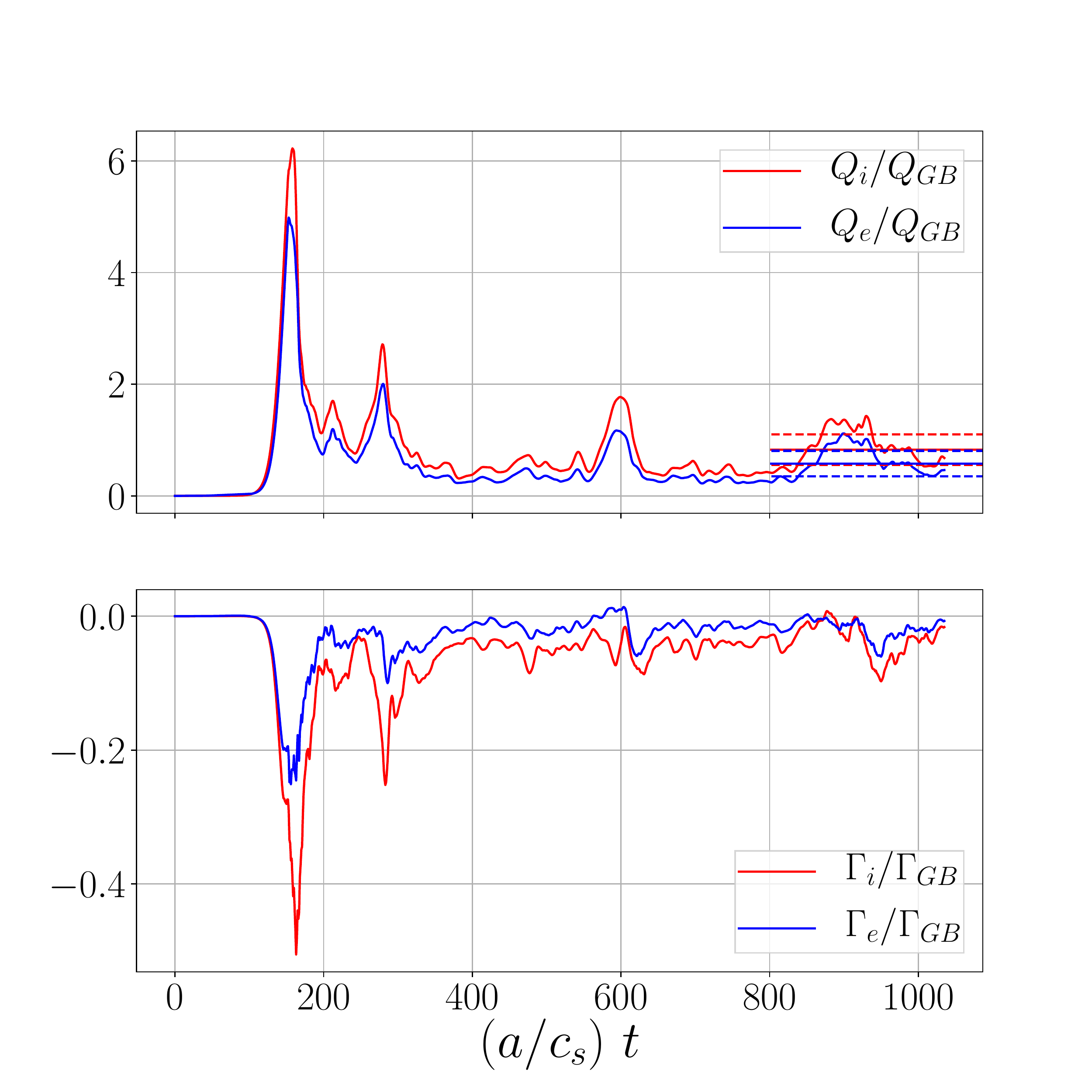}
  \caption{}
  \label{fig:imode_fluxes_3}
\end{subfigure}%
\begin{subfigure}{.5\textwidth}
  \centering
  \includegraphics[width=.99\linewidth]{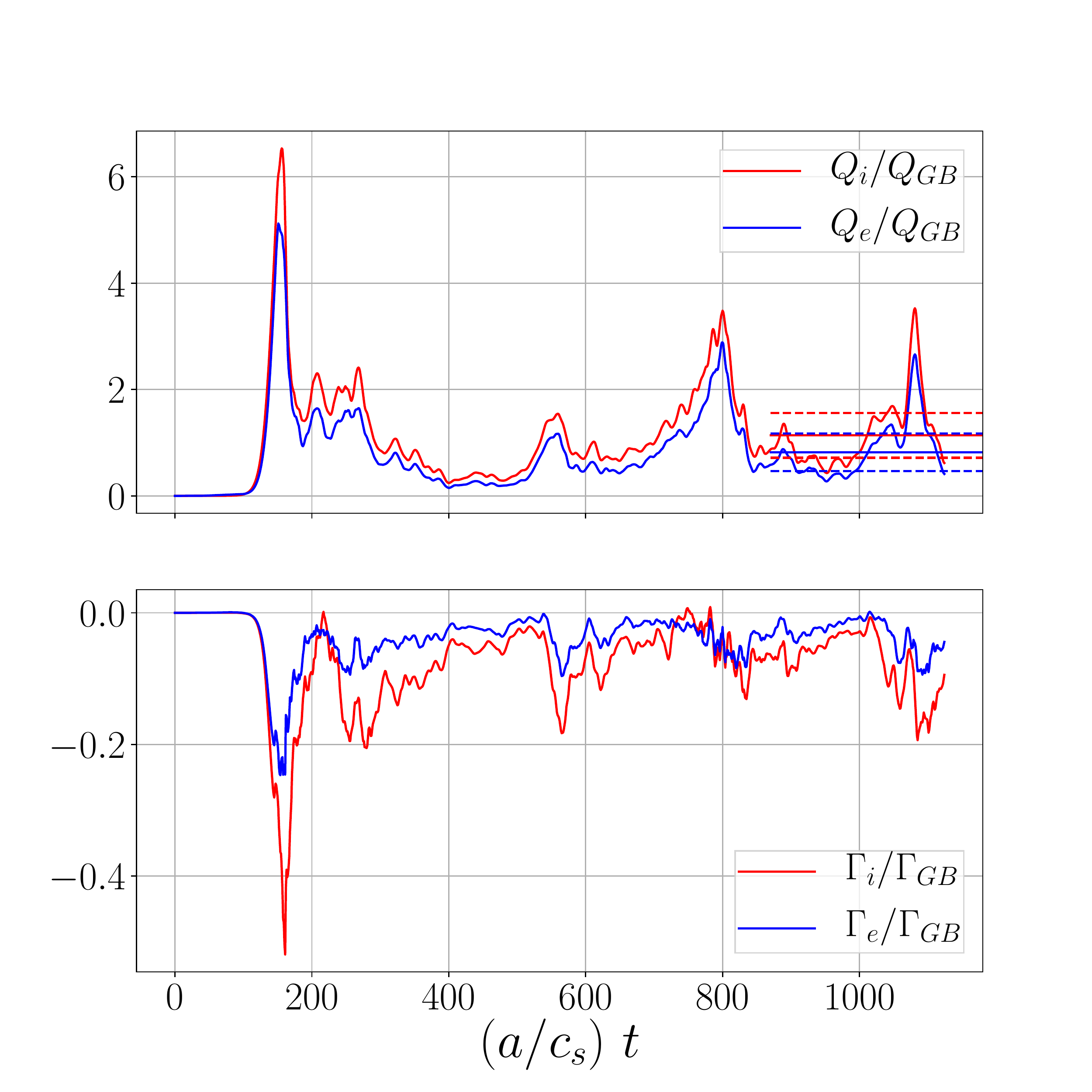}
  \caption{}
  \label{fig:imode_fluxes_4}
\end{subfigure}
\caption{Time evolution of total ion and electron fluxes in nearly $Q_i$-matched CGYRO simulations, differing in $a/L_{T_i}$ by only $0.5\%$. The case with larger $a/L_{T_i}$ (right) displays significantly more intermittency. We use a moving average of heat fluxes over the last 25\% of each simulation (dashed lines) to compare to experimental values.}
\label{fig:imode_fluxes}
\end{figure}

In Fig.~\ref{fig:imode_fluxes} we show the time history of heat and particle fluxes for two simulations that differed by only $0.5\%$ in their $a/L_{T_i}$ values. As a result of changing the temperature gradient by such small amount, turbulent intermittency is seen to vary significantly. This may be expected given the marginality and low heat fluxes in this regime. Horizontal lines in the heat flux plots (top) display the moving average over the last 25\% of the simulation, giving $Q_i^{sim}/Q_{gB}\approx 0.83\pm 0.27$ and $1.14\pm 0.47$ for the simulations with lower $a/L_{T_i}$ (left) and higher $a/L_{T_i}$ (right), respectively, where $Q_{gB}\coloneqq n_e T_e c_s (\rho_s/a)^2$ is the local gyro-Bohm unit of heat flux. 
Both of these values are reasonably close to the experimental estimate of $Q_i^{expt}/Q_{gB}\approx 1.09\pm0.3$, slightly under- or over-predicting it within uncertainty.

\begin{figure}[ht]
	\centering
	\includegraphics[width=0.7\textwidth]{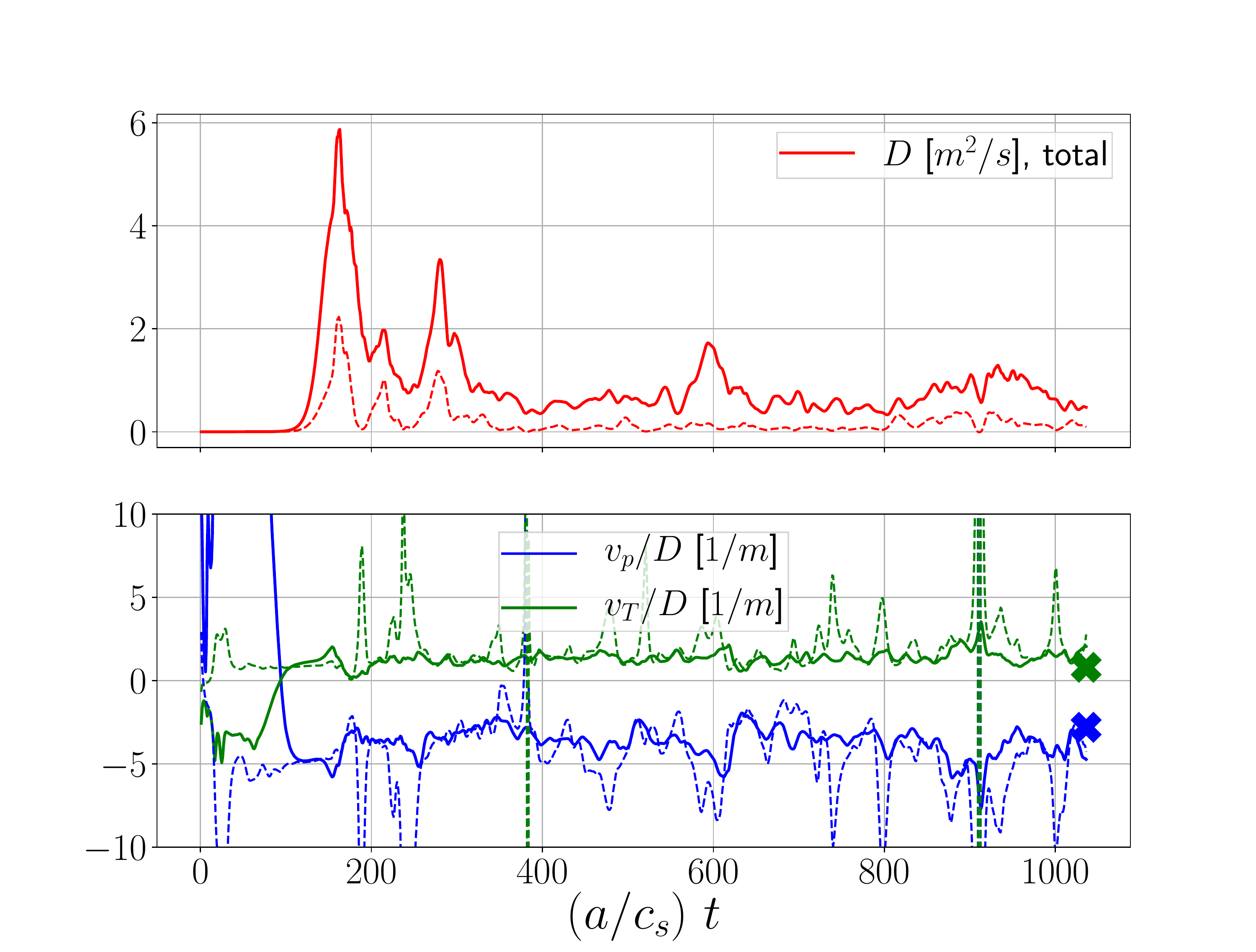}
	\caption{Time evolution of diffusive and convective transport components of trace impurity species over time for the lower-$a/L_{T_i}$ simulation of Fig.~\ref{fig:imode_fluxes_3}. Dashed lines indicate how transport coefficients computed only from $k_y \rho_s \approx 0.4$ track the total coefficients. Crosses on the right hand side of the lower panel show $v/D$ predictions from a linear CGYRO simulation at $k_y \rho_s = 0.4$.} 
	\label{fig:imode_DVtime}
\end{figure}

In Fig.~\ref{fig:imode_DVtime} we show the time evolution of Ca transport coefficients from the simulation in Fig.~\ref{fig:imode_fluxes_3}, again separating pure and thermal convection as in the TGLF simulations shown in section~\ref{sec:tglf}.
Continuous lines are the quantities obtained by the sum of all toroidal modes; dashed lines show results obtained from only the strongest mode, found via linear simulations to be at $k_y \rho_s\approx 0.4$. While the dashed line obviously does not match the total $D$ in magnitude (continuous line), $v/D$ is closely matched for each convection component separately; all discrepancies between single-mode and overall traces are due to transient growth of other toroidal modes. Furthermore, crosses at the end of the simulation time domain in the second panel indicate $v/D$ results obtained from a purely linear simulation at $k_y \rho_s = 0.4$. The proximity of these results suggests that quasilinear estimates (such as those obtained via TGLF, see Fig.~\ref{fig:DV_inference_w_models_042020}) should be able to closely match fully-nonlinear gyrokinetic predictions for $v/D$.
On the other hand, these results also show that matching $D$ and $v$ components separately between experiment and theory offers a much more stringent test than matching $v/D$ alone. We note that $v_T/D$, the thermal convection peaking component, is positive (as expected in the ITG regime, in agreement with TGLF) and it is smaller than the pure pinch; since $v_T$ scales with $1/Z$\cite{Angioni2012Off-diagonalAspects}, this may not be surprising for Ca impurities. 


\begin{figure}[ht]
	\centering
	\includegraphics[width=0.5\textwidth]{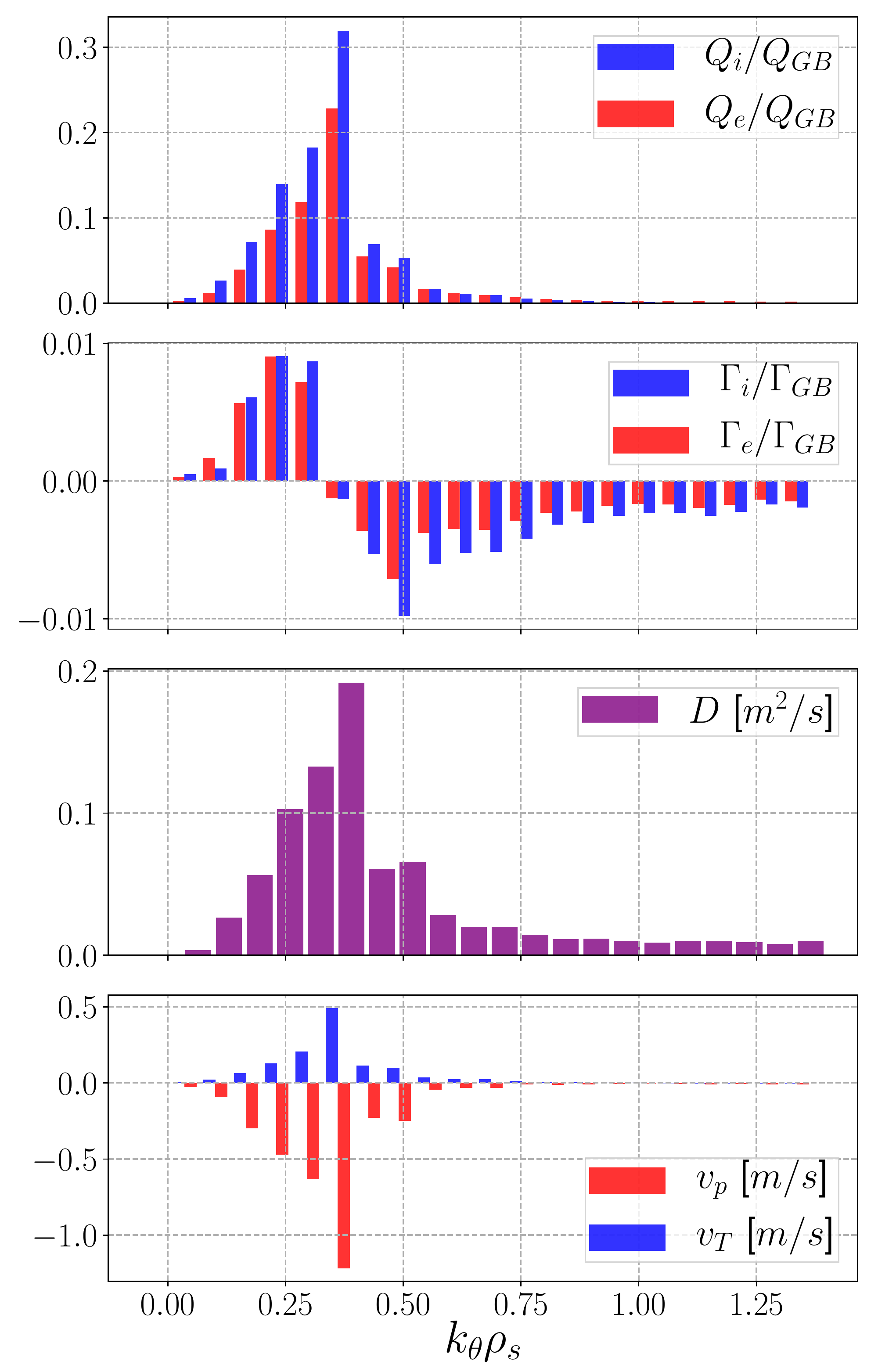} 
	\caption{Wavenumber spectra of heat and particle fluxes for both electrons and (total) ions, as well as of trace impurity diffusion ($D$) and convective components (pure, $v_p$, and thermal convection, $v_T$) over the non-linearly simulated range of $k_y \rho_s \lesssim 1.4$. These spectra correspond to the lower-$a/L_{T_i}$ case of Fig.~\ref{fig:imode_fluxes_3}.}
	\label{fig:Imode_bar_charts}
\end{figure}

In Fig.~\ref{fig:Imode_bar_charts} we show spectra of heat and particle fluxes, as well as diffusion and convective components of trace Ca impurities, again for the simulation corresponding to Fig.~\ref{fig:imode_fluxes_3}. Here, $\Gamma_{gB}\coloneqq  n_e c_s (\rho_s/a)^2$ is the gyro-Bohm unit of particle flux. 
Interestingly, the inversion of main-ion ($\Gamma_i$) and electron particle flux ($\Gamma_e$) at $k_y \rho_s\approx 0.35$ is not observed with the Lorentz collision operator, which predicts negative deuterium flux at all ion scales and larger positive electron particle flux at $k_y \rho_s < 0.4$. We note a clear similarity between heat flux spectrum and the impurity diffusion spectrum, consistently with the fact that heat transport is mostly of diffusive nature. Impurity convection terms do not seem to be significant for $k_y \rho_s\gtrsim 0.5$, i.e. at electron scales all transport appears to be of diffusive nature.

\section{Discussion and Summary}\label{sec:discussion}

In previous sections, we presented the results of experimental inferences of radial profiles of Ca transport coefficients and compared them to predictions from neoclassical and turbulent transport models. The nested sampling algorithm has allowed us to reliably estimate the Bayesian evidence and do model selection for appropriate and flexible parameterizations of transport coefficients and nuisance parameters. Doing so with an inference framework based on a minimization algorithm, such as the Levenberg-Marquart one, would have required questionable approximations and compromises. Ultimately, a comparison between algorithms would be valuable, but requires the construction of a database of inferences, which will be part of future work. 

The complexity of our modeling, both on the experimental and theoretical sides, leaves room for multiple potential sources of error in the validation process. From an experimental perspective, one may particularly expect inaccuracies due to diagnostic time bases synchronization or MHD activity in the plasma core. The use of a 1D model and the omission of thermal charge exchange in the plasma edge may also be problematic. While we cannot easily quantify uncertainties in ADAS effective ionization and recombination rates, we have thoroughly explored the sensitivity of our results to photon emissivity rates. Having obtained similar results with and without the inclusion of the Ca$^{18+}$ z-line, which depends on different atomic physics from the w-line, and having tested our inferences with and without time-dependent atomic rates (modulated by sawteeth in our experiment), we can confidently state that our results are robust to such issue. Since Ca is non-intrinsic and non-recycling, we have been able to avoid significant modeling complications related to SOL transport, thus making our results more robust. Inferred uncertainties have been found not to be strongly sensitive to prior shapes and hyperparameters, except for $v/D$, which is only weakly constrained by our data and is therefore effectively ``regularized'' by our prior. Increasing the $v/D$ prior width to larger values can indeed lead to inferences with higher peaking. While such solutions cannot be excluded due to our limited measurements, we note that our prior is consistent with expectations of $R/L_{n_z}$ not being dramatically different from $R/L_{n_e}$. 

We highlight that our experimental $D$, $v_{tot}$ uncertainties are constrained by the ``model'' that we apply, i.e. the number of spline knots, the choice of $D$, $v_{tot}$ parameterization, the sawtooth crash model, etc.. More physical (e.g., 2D) or more flexible models (not necessarily with more free parameters) would improve uncertainty quantification by overcoming the apparent model inadequacy, particularly in the pedestal. Still, even taking inferred uncertainties at face values, we note that TGLF scans (Fig.\ref{fig:Imode_DVscans_101318}) have shown sensitivity of modeled $D$, $v$ values to several inputs, making it plausible to have a closer match between experimental and modeling predictions than at first apparent from Fig.\ref{fig:DV_inference_w_models_042020}.

Interestingly, the application of a Z-dependence to the particle convection in the pedestal region was found to match experimental data better than a model with no Z-dependence for charge states, based on the Bayesian evidence. This observation has been corroborated with inferences on discharges that have not been described in this paper. In all cases, our experimental signals only offer weak constraints in the pedestal, therefore suggesting that even our core-focused measurements may constrain pedestal transport to some extent.

Modeling of the I-mode condition of interest with TGLF has shown that experimentally-inferred diffusion can be approximately matched within uncertainties, whereas convection at mid-radius is predicted to be smaller than experimentally inferred. Non-linear ion-scale $Q_i$-matched CGYRO simulations predicted smaller $D$,$v_{tot}$ values at $r/a=0.6$, thus increasing discrepancies with experiment with respect to TGLF. Although these simulations could not match experimental results quantitatively, we used them to elucidate the role of pure and thermal convection in determining peaking for our I-mode condition. Density peaking, being related to the $v/D$ ratio, has been found to be well described by the toroidal mode at $k_y=0.4$, both in linear and nonlinear simulations. However, separate comparison of $D$ and $v_{tot}$, rather than their ratio, clearly provides a much stronger validation constraint. In section~\ref{sec:cgyro}, we have remarked that different gyrokinetic collision operators can affect particle fluxes much more strongly than heat fluxes. This observation will be the subject of future work.  

In summary, this paper showed how \texttt{BITE} addresses many of the issues identified in previous work, particularly bridging the gap between experimental and synthetic data. \texttt{BITE}'s advances range from improved X-ray spectral fitting to inclusion of sawtooth temperature and density modulation, from faster iterative speed using \texttt{pySTRAHL} on a computing cluster to improved pedestal modeling. Bayesian model selection was used to avoid under- and over-fitting of experimental signals, allowing us to rigorously estimate diffusion and convection radial profiles. We compared our results to neoclassical NEO and quasilinear turbulent TGLF simulations across the radial extent, as well as to a single CGYRO nonlinear prediction at midradius, showing qualitative (and in some cases quantitative) agreement. The presented methods offer significant advances for data analysis on C-Mod and other devices, but are obviously no substitute for high quality experimental data. Future work will aim at further constraining transport inferences across the plasma, particularly focusing on expanding diagnostic capabilities and physical model fidelity.

\section*{Acknowledgments}
We are thankful to the Alcator C-Mod team for outstanding work over the years and to J. Candy, E. Belli, O. Meneghini and S. Smith for helpful conversations on transport modeling. Part of our analysis was performed using the OMFIT integrated modeling framework\cite{Meneghini2015IntegratedOMFIT}. This work was directly supported by US DoE grants DE-SC0014264 and made use of the MIT-PSFC partition of the \emph{engaging} cluster at the MGHPCC facility, funded by DoE award No. DE-FG02-91-ER54109, as well as the National Energy Research Scientific Computing Center (NERSC)–a US DoE Office of Science User Facility operated under Contract No. DE-AC02-05CH11231.

\begin{appendices}

\section{Bayesian Signal Combination}
\label{sec:appendixB}

The technique described in the main text to combine experimental datasets, e.g. from different plasma diagnostics, is an extension of methods that have been adopted in the astrophysical literature for many years. As in Ref.\cite{Hobson2002CombiningEvidence}, we look for the distribution of $\alpha_k$ of maximum entropy
\begin{equation*}
    S[P(\alpha)] = - \int_0^\infty P(\alpha_k) \ln P(\alpha_k) d\alpha_k
\end{equation*}
under the constraint of normalization $\int_0^\infty P(\alpha_k) d\alpha_k =1$ and unit mean $E[\alpha_k] = \int_0^\infty \alpha_k P(\alpha_k) =1$. This sets our expectation that the appropriate weight $\alpha_k$ for each diagnostic is 1. Such constrained optimization can be solved analytically using Lagrange multipliers and gives the exponential solution $P(\alpha_k) = \exp(-\alpha_k)$, which may be taken to appropriately represent our state of knowledge about $\alpha_k$. 

As pointed out in Ref. \cite{Hobson2002CombiningEvidence}, if we were interested in limiting the range of $\alpha_k$ values to be explored, one could follow the same maximum entropy recipe described above, but now adding an additional constraint on the variance of $\alpha_k$. This optimization cannot be analytically solved unless we extend the domain of  $\alpha_k$ to $[-\infty, +\infty]$, thus allowing for non-physical weights, to find the familiar Gaussian distribution. This domain extension can be problematic, and we address it by exploiting the fact that the exponential solution found above is a special case of the gamma distribution 
\begin{equation} \label{eq:gamma_distr}
    p(x|a,b) = \frac{1}{\Gamma(a) b^a} x^{a-1} e^{-x/b}
\end{equation}
for $a=1$,$b=1$. Generalizing the maximum-entropy argument above, we therefore adopt a gamma distribution that resembles a ``skewed Gaussian'' and still obeys the constraints above. 

\begin{figure}[ht]
	\centering
	\includegraphics[width=0.45\textwidth]{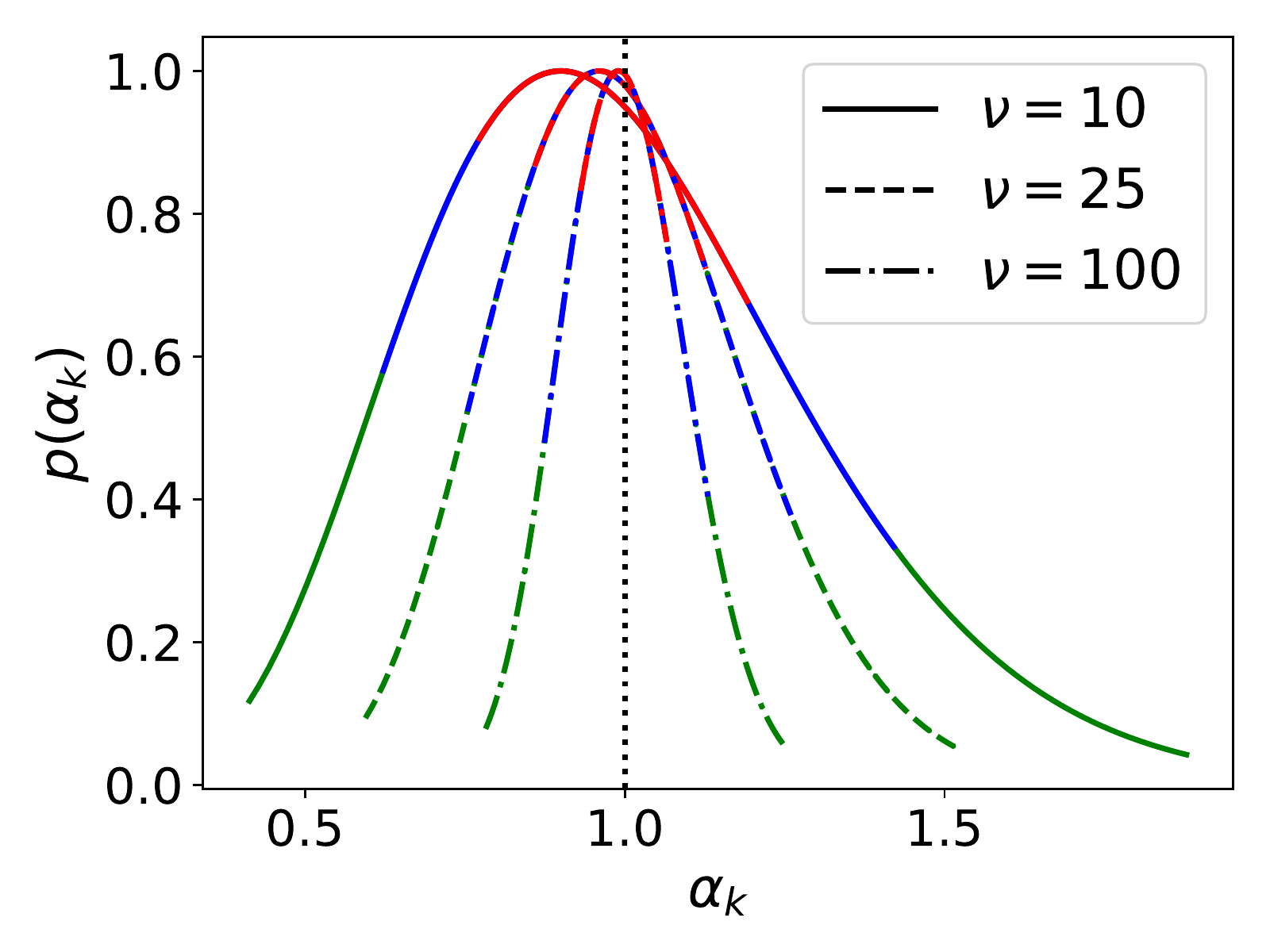}
	\caption{Gamma priors over diagnostic weights for some values of the $\nu$ parameter. All of these choices have a distribution mean of 1, indicated by the vertical black line. Quantiles of each distribution are shown in different colors: 25-75 (red), 10-90 (blue), 1-99 (green). $\nu=25$ was used in the inferences described in the main text.}
	\label{fig:alpha_k_priors}
\end{figure}

Integrating the product of the prior in Eq.~\ref{eq:gamma_distr} with a Gaussian likelihood, one has 
\begin{equation} \label{eq:prob_H2}
    P(D|\theta) = \prod_{k=1}^{N} \frac{1}{(2 \pi)^{n_k/2} |V_k|^{1/2} \Gamma(a) b^a} \int_0^{\infty} \alpha_k^{n_k/2} \  e^{-\frac{1}{2}\alpha_k \chi_k^2} \ e^{-\alpha/b} \  d\alpha
\end{equation}
where $V_k$ are the measurement covariance matrices, and $a$,$b$ are the gamma distribution parameters of Eq.~\ref{eq:gamma_distr}. After some algebra, and dropping constant factors that also appear in the standard Gaussian log-likelihood, this integration directly leads to Eq.~\ref{final3}. This allows one to impose finite variance for the prior over signal weights. Although the gamma distribution is not exactly a maximum-entropy solution, it is a logical extension to achieve our objectives. In order to fix the expectation value of the gamma prior to $1$, we set $\nu \vcentcolon= a=1/b$. Fig.~\ref{fig:alpha_k_priors} shows the resulting distribution for a number of choices of $\nu$.

One slight drawback of marginalizing over the weights prior is that no single value of the weights is directly inferred; rather, the entire prior distribution is indirectly explored. However, one may obtain an estimate for which weights were found to be most likely by differentiating Eq.~\ref{eq:prob_H2} with respect to $\alpha_k$ and setting the result to zero. This gives an ``effective'' weight for each diagnostic 
\begin{equation} \label{eq:eff_weight}
    \alpha_k^{eff} = \frac{(n_k + 2 a -2 ) b}{b \chi^2 +2}.
\end{equation}
The corresponding effective weights that would be obtained with an exponential prior are
\begin{equation} \label{alpha_eff}
    \alpha_k^{eff}(\theta) = \frac{n_k}{\chi_k^2(\theta)}
\end{equation}
Using Eq.~\ref{eq:eff_weight} in our inferences with $a=1/b=25$, we usually infer values in the interval $0.8<\alpha_k<1$ for XICS and $0.6<\alpha_k<0.8$ for XEUS. This suggests that wider experimental uncertainties are needed to obtain a Gaussian distribution of data points around the inferred solution, especially for the latter diagnostic, for which background subtraction and signal binning is expected to be less accurate. 

We note that, even without marginalizing over a weights prior, one could allow such weights to be inferred as free parameters. This requires weights to be included within the likelihood model, e.g. in the denominator of a Gaussian likelihood, and not only in the exponential. Of course, this increases the problem dimensionality and therefore also the runtime, making the analytic marginalization over signal weights described above more attractive.


\section{Application of \texttt{MultiNest}} \label{sec:Appendix_MultiNest}

As described in section~\ref{sec:ns}, nested sampling offers the means to quantify both the Bayesian evidence and the posterior distribution, offering an excellent alternative to Markov Chain Monte Carlo (MCMC) methods. \texttt{MultiNest} is an implementation of NS that fits ellipsoids to the set of live points and samples from within their union. We refer the reader to the Refs. \cite{Feroz2008MultimodalAnalyses,Feroz2013ImportanceAlgorithm} for details of the numerical methods. In this appendix, we describe only key features and choices of important parameters that we made for the present work. 

One attractive feature of \texttt{MultiNest} is its speed and effectiveness in exploring parameter spaces with up to approximately 30-50 dimensions. While other algorithms, such as \texttt{PolyChord}\cite{Handley2015POLYCHORD:Cosmology}, perform better at higher dimensionality, \texttt{MultiNest} is the ideal tool for our transport inferences extending to a maximum of 30 free parameters. For standard (non-dynamic) nested sampling, the larger the number of live points ($n_{live}$) used, the lower the chance of missing important parameter space. In our inferences, we vary the number of live points based on the inference dimensionality ($D$) in order to keep the evidence error constant\cite{Handley2015POLYCHORD:Cosmology}, using $n_{live}=200+25\times D$. 

\texttt{MultiNest} defines a \emph{target efficiency}, or ``inverse enlargement factor'', $f$, which expands the ellipsoids' hyper-volume ($0<f\leq1$) to avoid over-shrinking at any iteration, at the cost of slower convergence. In our inferences, we conservatively set $0.1$ when using vanilla NS and 0.01 for INS. Increasing $n_{live}$ or decreasing $f$ has been observed not to affect our results, while obviously incurring higher computational cost. 


In its INS variant, \texttt{MultiNest} can reach significantly faster convergence, particularly in its ``constant-efficiency'' mode. This makes INS a convenient choice for dimensionality scans, although \texttt{MultiNest}'s ability to isolate posterior modes is only available in vanilla NS. We make use of the latter to identify which experimental measurements and priors are better suited to exclude multi-modality from our impurity transport inferences. Such data-driven approach allows us to understand ambiguities of our data that may lead to unphysical local minima in the $D$ and $v$ posterior distributions.

\section{XEUS signal fits}
\label{sec:signal_fits}
\begin{figure}[ht]
	\centering
	\includegraphics[width=0.7\textwidth]{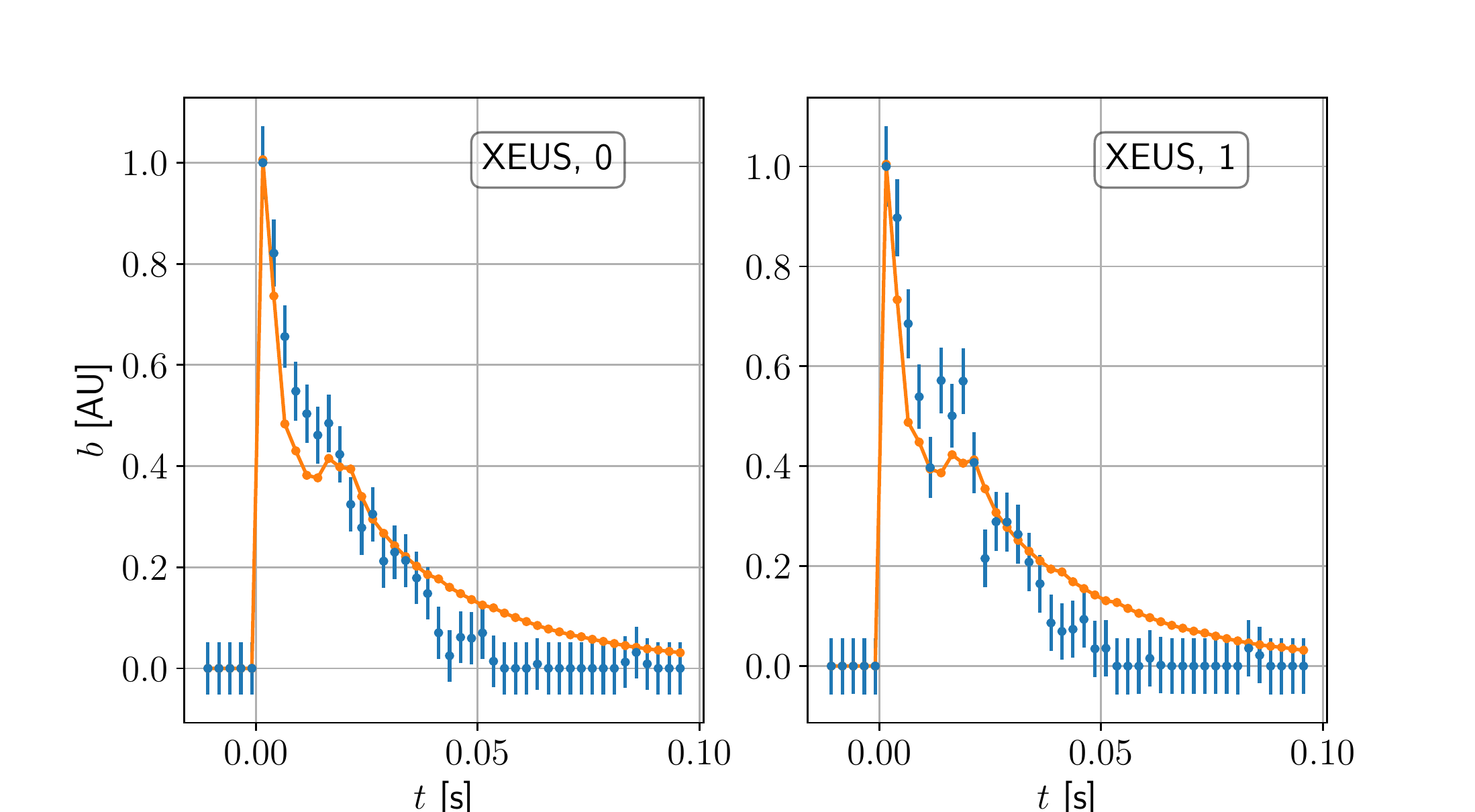}
	\caption{Signal fits for the XEUS groups of lines for the I-mode case discussed in the main body.}
	\label{fig:imode_xeus_all}
\end{figure}

For the sake of completeness, we show the XEUS spectroscopic signal matches from the I-mode case that was described in the main text. Sample signal fits for the XICS Ca w and z lines are shown in Fig.~\ref{fig:signals_decays}. Signal fits for the two VUV groups of lines measured by XEUS (viewing Ca$^{17+}$ emission) are shown in Fig.~\ref{fig:imode_xeus_all}. 



\end{appendices}
\clearpage

\bibliographystyle{vancouver} 
\bibliography{references,references_2}

\end{document}